\documentclass{aa}  
\usepackage{amsmath}
\usepackage{stfloats}

\usepackage{graphicx}
\usepackage{txfonts}
\usepackage{lipsum}
\usepackage[colorlinks=true,linkcolor=red, citecolor=blue]{hyperref}
\usepackage{etoolbox}

\usepackage{enumitem}
\usepackage{subcaption}         
\usepackage{lscape}             
\usepackage{placeins}           
                                
\newcommand{\modelitem}[2]{%
  \item \hypertarget{#1}{}\textbf{#2}\label{#1}\\
}

\usepackage{comment}
\begin{document}

    \titlerunning{Constraining models using photometric neighbors at $z>10$}
   \title{Constraining the galaxy-halo connection at $z>10$ \\using the neighbors of UV-bright galaxies}


   \author{Ivan Nikolić \inst{1,2}
        \and Charlotte A. Mason \inst{1,2}
        }
    \institute{
        Cosmic Dawn Center (DAWN), Denmark
        \and
         Niels Bohr Institute, University of Copenhagen, Jagtvej 128, 2200 Copenhagen N, Denmark
    }

 
  \abstract
   {JWST has discovered an overabundance of UV-luminous galaxies compared to extrapolations from pre-JWST galaxy formation models.
   A promising path to distinguish between proposed theoretical explanations requires constraining the halo masses of these systems, which can be probed through clustering. 
   However, at $z>10$, low galaxy number densities render traditional clustering measurements challenging. 
   We introduce a novel formalism that uses information about the local environments ($\lesssim$2 arcmin) of individual UV-bright galaxies at $z>10$ to test the galaxy-halo connection implied by various theoretical models for the UVLF excess. We create mock JWST observations using (500\,cMpc)$^3$ 21cmFAST simulations to forward-model the spatial distribution of fainter photometric candidates around UV-bright galaxies. 
   We find that the distance to the nearest neighbor is sensitive to the stochasticity of the UV luminosity-halo mass relation, and can distinguish between models where UV luminosity is boosted uniformly in galaxies versus models with high stochasticity in UV luminosity at fixed halo mass.
   As a proof-of-concept, we apply our framework to observations around the two brightest confirmed galaxies at $z>10$, GN-z11 and JADES-GS-z14-0, and find that the lack of neighbors within $<0.5$ arcmin of these sources disfavors models where UV luminosity is boosted uniformly, suggesting that stochasticity is important for explaining their high UV luminosities.
   We forecast that distinguishing between these models at high significance requires deep imaging surveying $<1.1$\,arcmin around $7$\,--\,$15$ bright galaxies ($M_{\rm UV}\lesssim-21$) at $z =10$\,--\,$12$,  dropping to $\lesssim5$ at $z\sim14$, capable of detecting their neighbors down to $M_{\rm UV,lim}\approx-18$.
   These observations are within reach of JWST, Roman and Euclid surveys, providing a path to constraining the galaxy-halo connection at our redshift frontier.

   }


   \maketitle

\section{Introduction}

JWST has revolutionized our view of galaxy evolution at the earliest epochs \citep{Gardner2006, Rigby2023}, with deep extragalactic surveys \citep[e.g.,][]{Eisenstein2026, Finkelstein2025, Casey2023, Bezanson2024} pushing spectroscopic confirmations to $z>14$ \citep{Carniani2024, Naidu2025}. Observations of this previously inaccessible epoch have uncovered a higher-than-expected number density of luminous galaxies at $z>10$ \citep[quantified using the UV luminosity function (UVLF);][]{Austin2023, Castellano2023, Bouwens2023, Harikane2023, Finkelstein2023, Perez-Gonzalez2023, Morishita2023, Donnan2024, McLeod2024, Willot2024, Adams2024, Whitler2025, McLeod2026}, exceeding predictions from models calibrated on HST data that follow the steep expected decline in the dark matter halo mass function at these redshifts \citep[e.g.,][]{Mason2015,Tacchella2018,Yung2019}.

The physical origin of this excess remains unclear, but may imply a shift in the galaxy-halo connection at early times.
These observations have sparked a plethora of theoretical explanations, which essentially modify the expected distribution of galaxy properties for a dark matter halo of a given mass.
Broadly, these models fall into one of two classes: either early galaxies were on average brighter at fixed halo mass, due to e.g., increased star-formation efficiency \citep{Mason2023, Pallottini2024,  Nikopoulos2024, Somerville2025, Kar2026}, a top-heavy initial mass function \citep{Cueto2024, Jeong2025, Hutter2025}, lower dust attenuation \citep{Ferrara2023, Ziparo2023, Nakazato2025,Narayanan2026}, or enhanced AGN activity \citep{Pacucci2022, Hegde2024}; or, high stochasticity in star formation temporarily scatters galaxies in numerous low mass halos to bright luminosities, boosting the number of luminous galaxies \citep{Mason2023, Mirocha2023, Sun2023, Shen2023}. It has been suggested that part of the $z>10$ evolution may be due to the dependence on halo mass of a redshift-independent star formation efficiency and/or stochastic SFH \citep{Gelli2024,Feldmann2025,Munoz2026}, but this may not be sufficient to explain the UVLF at $z\sim14$ \citep[][Gelli et al. in prep]{Munoz2026}.
Unfortunately, discriminating between these physical scenarios is challenging using the UVLF alone \citep[e.g.,][]{Mirocha2020,Munoz2023}, as, by construction, these new models tend to all match the UVLF over the observed magnitude and redshift range, despite predicting very different galaxy-halo connections.

Measurements of galaxy clustering provide a potential key to isolate the primary cause of the UVLF excess \citep[e.g.,][]{Munoz2023,Gelli2024}.
Since halos form hierarchically, more massive halos cluster more strongly \citep{Mo1996}. Galaxies, as tracers of dark-matter halos, inherit this clustering \citep{White1978}, allowing us to probe the halo population hosting the observed galaxies.
Clustering is most commonly quantified through the angular correlation function, which measures the excess probability of finding galaxy pairs relative to a random distribution \citep{Peebles1973, Groth1977}. 
Recent JWST observations in extra-galactic legacy fields have enabled the first estimates of angular correlation functions up to $z\sim6-10$ \citep{Paquereau2025, Shuntov2025, Dalmasso2024, Dalmasso2026}.
JWST's pure-parallel imaging surveys PANORAMIC \citep{Williams2025} and BEACON \citep{Morishita2025} have also enabled angular clustering to be probed at $z\sim7-10$ by estimating variations of number densities of galaxies across independent pointings \citep{Weibel2025,Kreilgaard2026}. 
These studies have generally inferred that the most UV luminous galaxies are the most clustered, and thus likely trace massive halos at $z\lesssim10$. 
If the $z>10$ UVLF excess is mostly due to enhanced UV luminosity in galaxies at fixed halo mass, we expect the brightest galaxies to remain strongly clustered.

However, our understanding of galaxy clustering at the crucial redshift frontier $z>10$ is currently limited. At these redshifts, decreasing galaxy number densities make measurements of the angular correlation function challenging due to large Poisson uncertainties \citep{Endsley2020,Paquereau2025, Dalmasso2026}.
Progress requires using alternative estimates of galaxy clustering, which can overcome these uncertainties.
Intriguingly, current observations of the brightest $z>10$ galaxies suggest they may be in rich environments, hinting at strong clustering.
In particular, in fields with deep NIRCam imaging, many photometric neighbors have been reported near the brightest known $z>10$ galaxies.
The most famous example, GN-z11, \citep[$z\approx10.6, M_{\rm UV}\approx -21.5$, ][]{Oesch2016, Bunker2023}, has $9$ photometric neighbors within a $\sim 2$ arcmin ($\sim 6$ cMpc projected) radius \citep{Tacchella2023}. 
Similarly, a large number of photometric candidates have been reported around GS-z14-0 \citep[$z\approx14.2, M_{\rm UV} = -20.8$,][]{Carniani2024,Robertson2023,Whitler2025,Hainline2026}.
Other bright galaxies, such as GHZ2 \citep{Castellano2024, Napolitano2025} and MoM-z14 \citep{Naidu2025}, also provide promising future targets for environmental studies, but currently only have shallow and incomplete areal coverage.
These observations motivate new approaches to estimating galaxy clustering at $z>10$.

In this work, we introduce a framework to use information about the photometric neighbors around bright galaxies at $z>10$ to distinguish between galaxy models which have degenerate effects on the UVLF. We compare two representative models which reproduce $z>10$ UVLFs: one with a tight UV luminosity-halo mass relation, with enhanced luminosity relative to pre-JWST predictions, and one with a high amplitude of halo mass-dependent stochasticity in UV luminosity. To build intuition, we develop an analytic formalism to explore the predicted abundances and positions of bright galaxies' neighbors in these two models. We then use semi-numerical simulations to generate realistic forward-models of JWST observations, allowing us to construct a new summary statistic, the separation between the bright galaxy and its closest neighbor, that probes clustering at redshifts not previously accessible with standard methods. As a proof-of-concept, we apply this formalism to two of the brightest $z>10$ galaxies, GN-z11 and GS-z14-0, and show their environments suggest stochasticity is important in explaining their high UV luminosities.

This paper is structured as follows. Section~\ref{sec:analytical_framework} presents an analytical framework predicting the abundances and positions of neighbors around bright galaxies.  We increase the realism in Section~\ref{sec:methodology} by employing semi-numerical simulations for a more accurate treatment of 3D galaxy distributions, and describe our new separation summary statistic for estimating clustering. In Section~\ref{sec:obs} we apply this framework to GN-z11 and GS-z14-0. We discuss our results in Section~\ref{sec:discussion} and present our conclusions in Section~\ref{sec:conclusions}. We adopt the Planck 2018 reference cosmology \citep{Planck2020} and the AB magnitude system throughout.

\section{Connecting galaxies to their dark matter halos}
\label{sec:analytical_framework}

\begin{figure*}[h!]
    \centering
    \includegraphics[width=\linewidth]{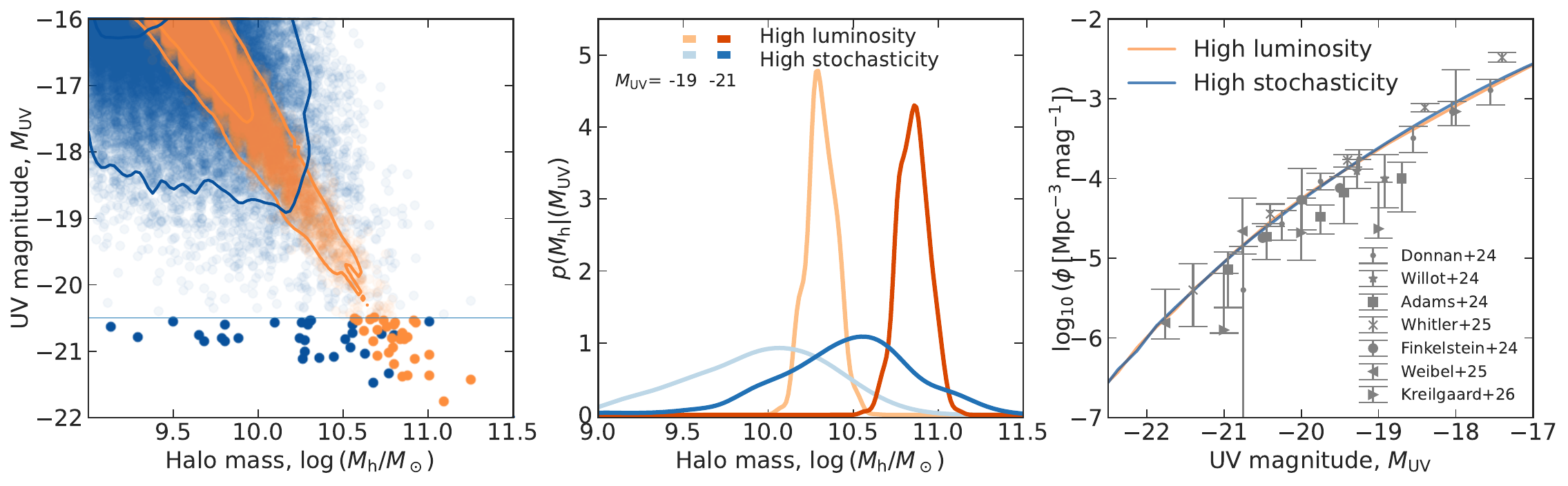}
    \caption{\textit{Left panel:} Joint distribution of galaxies in the $M_{\rm UV} - M_{\rm h}$ plane for two galaxy formation models, one in which the UV magnitudes of all galaxies are intrinsically brighter (orange line and points), and one in which stochasticity in the $M_{\rm UV} - M_{\rm h}$ connection is significantly increased (blue line and points). The shaded points are taken as samples from the joint distribution $p(M_{\rm UV}, M_{\rm h})$, while contours represent $68\%$ and $95\%$ of the samples for the two models. Also highlighted are galaxies with $M_{\rm UV}<-20.5$. \textit{Middle panel:} Conditional distributions of halo masses for two different $M_{\rm UV}$ values for the two models. Even though the two models reproduce almost identical abundance of galaxies for each $M_{\rm UV}$, they occupy different halo mass distributions. \textit{Right panel:} UV luminosity function at $z=10.5$ for the two galaxy formation models. The models are tuned to match the UVLF observations from JWST across different surveys \citep{Donnan2024, Adams2024, Willot2024, Finkelstein2023, Whitler2025, Weibel2025, Kreilgaard2026}. The two models cannot be distinguished based on the UVLF alone.}
    \label{fig:pmuv}
\end{figure*}

Galaxy properties correlate with the mass of their dark matter halos \citep[\citealt{White1978}, see review by][]{Wechsler2018}. However, the relation between the galaxy properties and the halo mass is not deterministic due to stochastic processes such as gas accretion, mergers, and feedback \citep[][]{McKee2007, Weisz2012, Agertz2015, Hayward2017, Faucher-Giguere2018, Matthee2019, vanLoon2021}. As such, galaxy properties are better described using conditional distribution functions. For example, the UV magnitude of a galaxy can be linked to halo mass via $p(M_{\rm UV}\,|\,M_{\rm h})$ \citep{Yang2003, Cacciato2009}. 

Modifications of this distribution relative to pre-JWST expectations were invoked to explain the overabundance of bright photometric candidates detected by JWST. Those modifications can be implemented either by increasing the mean relation \citep[increased efficiency, top-heavy initial mass function, lower dust attenuation;  see][]{Mason2023, Ferrara2023, Dekel2023,  Ziparo2023, Shen2024, Pallottini2024,  Nikopoulos2024, Cueto2024, Jeong2025, Hutter2025, Nakazato2025, Narayanan2026, Kar2026}, or increasing the width of the distribution \citep[increased stochasticity models; see][see also \citealt{Nikolic2024}]{Mirocha2023, Mason2023, Munoz2023, Munoz2026, Gelli2024, Sun2023, Kravtsov2024}\footnote{Models which consider changes in cosmology to explain the overabundance of bright galaxies alter the distribution of dark matter halos, but do not necessarily alter the galaxy-halo connection. We do not consider changes to the halo mass function in this work, but note that probing the clustering can also constrain cosmological parameters \citep[][]{Sailer2021}.}. Both classes of models reproduce the UV luminosity functions at various redshifts, but with a different galaxy-halo connection. In this work, we explore how to efficiently break the degeneracy between the two classes of models at $z>10$ using this information. We start by constructing analytical models of the galaxy-halo connection and mapping them to observed UV luminosity function.

\subsection{Analytical models of galaxy-halo connection}

We model the UV luminosity-halo mass relation, $p(M_{\rm UV}\,|\,M_{\rm h})$, as a normal distribution in magnitude space. This is a reasonable approximation, motivated by the log-normal nature of fluctuations of star-formation rate and the central limit theorem \citep[see, e.g.,][]{Tacchella2020, Munoz2023, Shen2023}. In shorthand, we write this as $p(M_{\rm UV}\,|\,M_{\rm h}) = \mathcal{N}\left(\mu_{\rm UV}(M_{\rm h}), \sigma_{\rm UV}(M_{\rm h})\right)$. Potential deviations from Gaussianity in the tails \citep[][Gelli et al., in prep]{Sun2025} should not significantly affect our results since the clustering statistics that we develop in this work are mostly sensitive to the mean (i.e., average galaxy-halo connection) and the scatter around it (which sets the amount of upscattering toward the bright end) \footnote{Non-Gaussian tails in $p(M_{\rm UV} | M_{\rm h})$ can have an effect on the bright population of galaxies, leading to them residing in lower mass halos than predicted in our fiducial stochastic model. However, this would only strengthen the preference for high stochasticity presented in Sec.~\ref{sec:obs}.}.

In order to study the galaxy-halo connection the two classes of models imply, we consider two representative models. In both cases we assume the median $\mu_{\rm UV}(M_{\rm h})$ relation derived by \citet{Mason2015,Mason2023}, based on abundance matching and halo accretion rates, which did not include scatter, to capture the shape of the UV luminosity-halo relation. This allows us to isolate the effects of an overall increase in UV luminosity, as well as stochasticity around the median relation, on the resulting mapping between galaxies and dark matter halos \citep{Mason2015, Behroozi2019, Tacchella2020}. 
To account for the impact of stochasticity and redshift evolution at $z\gtrsim10$, we modify $p(M_{\rm UV}\,|\,M_{\rm h})$, using a constant offset from the median, $\mu_{\rm UV, shift}$ (to account for a bulk increase in UV luminosity at fixed halo mass), and standard deviation, $\sigma_{\rm UV}$, of the UV luminosity-halo relation to capture the scatter, calibrated to reproduce $z\gtrsim10$ UVLFs. At $z\approx10$ we use the following parameters, chosen to broadly reproduce observed UVLFs \citep{Donnan2024, Willot2024, Adams2024, Whitler2025, Finkelstein2023, Weibel2025, Kreilgaard2026}:

\begin{itemize}

\modelitem{item:ib}{High luminosity model}
This model represents the limiting case in which stochasticity is minimized and luminosity is enhanced uniformly in galaxies to match the UVLF. 
We introduce a constant $\sigma_{\rm UV}=0.3$\,mag. This is the minimum scatter expected based solely on the stochasticity in accretion rates \citep[see e.g.,][]{Behroozi2010, Mutch2016, Ren2018, Ren2019}.
To match the normalization of the UVLF we increase the median $M_\mathrm{UV}(M_h)$ relation by $\mu_{\rm UV, shift} = -0.8$\,mag\footnote{While typically introducing stochasticity requires shifting the median luminosity to lower values to account for the increase in galaxies at the bright end \citep{Ren2019}, we find we still need to boost the luminosities to match JWST UVLFs at $z\approx10$ \citep{Donnan2024, Finkelstein2023, Weibel2025}}.

\modelitem{item:is}{High stochasticity model}
This model assumes a high amplitude of mass-dependent stochasticity, with the scatter normalized such that $\sigma_{\rm UV}$ reaches $\sim2$\,mag in atomic-cooling halos (i.e., halos with mass $M_{\rm h}\gtrsim10^{8.5}M_{\odot}$), corresponding to the largest values proposed in the literature \citep{Munoz2023, Shen2023, Munoz2026}.
We assume a mass-dependent scatter: $\sigma_{\rm UV} (M_{\rm h}) [\rm mag]= -0.34 (\log(M_{\rm h}/M_{\odot}) - 12) + 0.6$, with slope fixed to match the results of \citet{Gelli2024}, motivated by the high level of scatter seen in the FIRE simulations \citep{Sun2023}.
Given the high level of stochasticity, to match the normalization of the observed UVLF we decrease the median $M_\mathrm{UV}(M_h)$ relation by $\mu_{\rm UV, shift} = +0.3$\,mag \citep{Ren2019}.
\end{itemize}


We show the joint distribution of UV magnitudes and halo masses for the two models in the left panel of Fig.~\ref{fig:pmuv} obtained by sampling the $p(M_{\rm UV}\,|\,M_{\rm h})$ for halos sampled by the halo mass function \citep{Sheth2001}.
The two joint distributions of halo masses and UV magnitudes differ substantially, yet they reproduce the same number density of bright galaxies (e.g., those brighter than $M_{\rm UV} = -20.5$, highlighted with bold points), albeit hosted by halos of different masses. We further illustrate the differences between the models by showing the conditional distribution $p(M_{\rm h}\,|\,M_{\rm UV})$\footnote{Throughout this work we assume the minimum halo mass that harbor star-forming galaxies of $M_{\rm h,min}=10^{8.5}M_{\odot}$, corresponding to atomic cooling halos \citep{Barkana2001, Sobacchi2013}} for $M_{\rm UV} \in \{-19, -21\}$ in the middle panel of Fig.~\ref{fig:pmuv}. Galaxies of the same $M_{\rm UV}$ have vastly different halo mass distributions in the two models, with high stochasticity encompassing $\sim2$ dex in halo mass distribution for fixed $M_{\rm UV}$, compared to $\sim 0.5$ dex for the high luminosity model.

From the conditional distributions, we can also calculate the luminosity function (see also Appendix~\ref{sec:appB}, where we detail how UVLF is constructed from the halo mass function):

\begin{equation}
    \frac{\textrm{d}n}{\textrm{d}M_{\rm UV}} (M_{\rm UV}) = \int \textrm{d}M_{\rm h} \frac{\textrm{d}n}{\textrm{d}M_{\rm h}} p(M_{\rm UV}\,|\,M_{\rm h}),
    \label{eq:standard_uvlf}
\end{equation}
where $\frac{\textrm{d}n}{\textrm{d}M_{\rm h}}$ is the halo mass function, which we take from \citet{Sheth2001}. 
We plot the UVLF at $z=10.5$ for the two models in the right panel of Fig.~\ref{fig:pmuv}. The UV luminosity functions are completely degenerate and consistent with the JWST data \citep{Donnan2024, Adams2024, Finkelstein2023, Willot2024, Whitler2025, Weibel2025, Kreilgaard2026} by construction.

Clearly, the UV luminosity function alone is insufficient to disentangle these models. In contrast, because the two models trace different halos at fixed $M_{\rm UV}$, we expect their clustering signals to differ, with the high luminosity model expected to exhibit stronger clustering due to larger halo masses. This allows clustering statistics, such as the angular correlation function, to be used to break the degeneracy between the models \citep{Munoz2023}. However, measurements of the angular correlation function at $z>10$ remain highly uncertain with current JWST samples \citep{Paquereau2025, Dalmasso2024,Dalmasso2026}.

However, full two-point correlation function measurements are not required to probe galaxy clustering.  Instead, we develop summary statistics designed to capture the environments of bright galaxies, enabling us to assess whether galaxy observations are better explained by a model with enhanced UV emission or by increased stochasticity. We begin by building physical intuition using theoretical expectations from structure formation.

\subsection{Excursion set expectations for environments of bright galaxies}
\label{sec:analytic}
The number density of galaxies is spatially modulated by large-scale overdensities. To quantify this effect, we use the conditional mass function formalism \citep{Press1974, Bond1991}.
This formalism allows us to estimate the overdensity of a field centered on a UV bright galaxy from the abundance and spatial distribution of its neighboring galaxies. 
We present the main equations and results here and provide a full discussion in Appendix~\ref{sec:appB}.

The matter density of a region is not directly known, but we can infer it using the observed galaxy population \citep[see also][]{Trapp2022, Trapp2023}. In particular, the presence of a UV bright galaxy is expected to be a good indicator of an overdense region \citep{Munoz2008, Trenti2012, Lim2025}. This means we can relate the expected number of galaxies in a volume $V$ to the UV luminosity of the brightest galaxy in that region $M_{\rm UV, 0}$:

\begin{equation}
\begin{split}
N_{\rm gal}(\vec{r}, V|M_{\rm UV, 0})  & = \int \textrm{d}\delta \,N_{\rm gal}(V\,|\,\delta)\, p(\delta\,|\,M_{\rm UV,0}),
    \label{eq:n_gal_conditioned}
\end{split}
\end{equation}

where $N_{\rm gal}(\vec{r}, V|M_{\rm UV, 0})$ is the expected number of galaxies in a region of volume $V$ at spatial position $\vec{r}$, conditioned on the presence of a bright galaxy with magnitude $M_{\rm UV,0}$. 
The quantity $\delta$ represents the overdensity of the region, defined as $1+\delta = \frac{\rho_m}{\overline{\rho}_m(z)}$, where $\overline{\rho}_m(z)$ is the mean matter density of the Universe at a given redshift. Finally, $N_{\rm gal}(V|\delta)$ is the expected number of galaxies, above a magnitude limit $M_{\rm UV,lim}$, with overdensity $\delta$ in a volume $V$ (see Appendix~\ref{sec:cmf} for details on calculating this number). The term $p(\delta\,|\,M_{\rm UV,0})$ is the probability density function of the overdensity $\delta$, conditioned on the presence of a galaxy with UV magnitude $M_{\rm UV,0}$. This probability distribution can be obtained from Bayes' theorem and the galaxy-halo connection encoded in $p(M_{\rm UV}\,|\,M_{\rm h})$ (see Appendix~\ref{sec:cdd} for derivation of this quantity).

\begin{figure*}
    \centering
    \includegraphics[width=\linewidth]{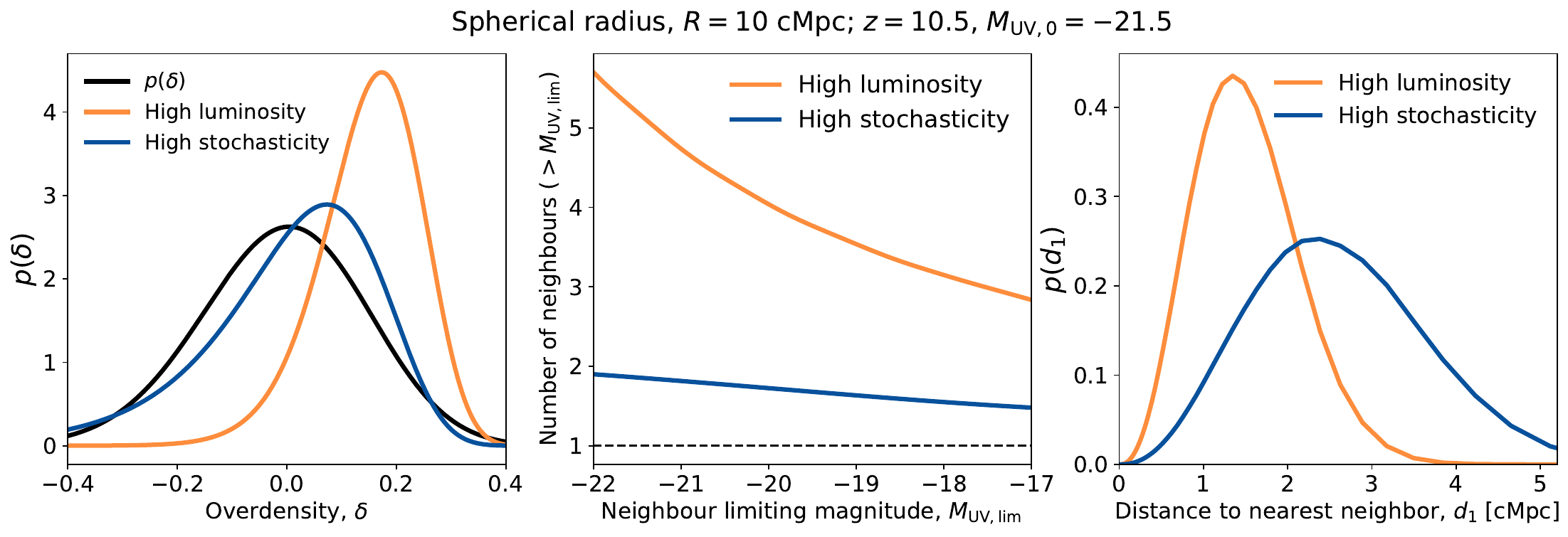}
    \caption{Theoretical predictions of environments of bright galaxies. \textit{Left panel:} Overdensity distribution of a spherical region of radius $R=10$ cMpc at $z=10.5$, conditioned on the presence of a galaxy with $M_{\rm UV,0} = -21.5$, for the high luminosity (orange) and high stochasticity (blue) models. The distribution of overdensities of a random region is shown in black. Bright galaxies are more likely to be in dense regions, with more massive halos predicted in the high luminosity model shifting the distribution to higher overdensities compared to the high stochasticity model. \textit{Middle panel:} Fractional increase in the number of galaxies in a spherical region of $R=10$ cMpc at $z=10.5$, conditioned on the presence of a bright galaxy with $M_{\rm UV,0} = -21.5$, for the high luminosity (orange) and high stochasticity (blue) models. The enhanced clustering in the high luminosity model implies a larger increase in the number density of galaxies around a bright object. \textit{Right panel:} The radial distance to the nearest neighbor brighter than $M_{\rm UV,lim} = -18.5$ for the high luminosity (orange) and high stochasticity (blue) models. Enhanced clustering in the high luminosity model implies shorter distances to neighboring galaxies.}
    \label{fig:full_theory_plot}
\end{figure*}

We visualize the distribution of overdensity given the presence of a UV bright galaxy, $p(\delta\,|\,M_{\rm UV,0})$, to gain intuition about the environments of bright galaxies. 
In the left panel of Fig.~\ref{fig:full_theory_plot}, we show the prior distribution of overdensities, $p(\delta)$ for a spherical region of size $R_{\rm b} = 10$ cMpc at $z=10.5$ in black (where the size and redshift roughly correspond to observations of GN-z11 from \citet{Tacchella2023}), along with the overdensity distribution for the same size region, but centered on a UV bright galaxy ($M_{\rm UV, 0} = -21.5$) in both the high luminosity and high stochasticity models (colored lines). 
\textit{Randomly} selected fields, not centered on a bright galaxy, would sample the prior distribution $p(\delta)$, a Gaussian random distribution of the dark matter density field. However, bright galaxies (in both models) are more likely to be hosted in relatively massive halos ($M_h\sim10^{10}-10^{11}\,M_\odot$ at $z\sim10$, see Fig.~\ref{fig:pmuv}), which are biased tracers of the density field.
As a result, the overdensity distribution around bright galaxies, $p(\delta\,|\,M_{\rm UV,0})$, is shifted to higher values than the overdensity distribution for random fields, in both models.
Fig.~\ref{fig:full_theory_plot} also shows that in the high luminosity model, a galaxy of fixed $M_{\rm UV}$ is more likely to reside in a stronger overdensity than in the high stochasticity model (peaking at $\delta \sim 0.2$ compared to $\delta \sim 0.1$). Additionally, the range of overdensities a bright galaxy resides in is narrower in the high luminosity model, compared to the high stochasticity model. 
This is consistent with the UV luminosity-halo mass relation in the two models, as shown in Fig.~\ref{fig:pmuv}. In the high luminosity model, a UV bright galaxy is likely to reside in a high mass halo, with a narrow mass range, thus the most biased regions of the density field. Meanwhile, in the high stochasticity model the same luminosity galaxy is on average hosted in a lower mass halo, and can be hosted in a wider range of halo masses, thus sampling a broader range of the density field.

The dark matter structure of the field impacts halo formation on all mass scales. For example, an over-dense region contains a higher number of dark matter halos at all masses \citep{Bond1991}. As a consequence, the presence of an overdensity around a bright galaxy is expected to increase the number of fainter (and potentially unobserved) galaxies in the field.
We calculate the expected number of galaxies around a bright galaxy with UV magnitude $M_{\rm UV,0}$ given some $M_{\rm UV}$ limit (see Appendix~\ref{sec:final_expression} for the final form of the equation). To quantify the impact of an elevated density around the bright galaxy, we compare this to the expected number of galaxies in a random field of the same size. We do this by calculating:

\begin{equation}
\begin{split}
    \Delta N (>M_{\rm UV,lim} & | M_{\rm UV,0} = -21.5)  \equiv \\ & \frac{N_{\rm gal}(>M_{\rm UV,lim}|M_{\rm UV,0} = -21.5) ~ [\rm from\ Eq.~\ref{eq:n_gal_conditioned}]}{N_{\rm gal}(>M_{\rm UV,lim}) ~ [\rm from\ the\ integral\ of\ Eq.~\ref{eq:standard_uvlf}]}
\end{split}
\end{equation}

We show this quantity in the middle panel of Fig.~\ref{fig:full_theory_plot} for the two models as a function of $M_{\rm UV}$ of neighbors at $z=10.5$ for a spherical volume of $R=10$ cMpc. Both models show an increase in the expected number of galaxies relative to the expectation in a mean-density region.
The high luminosity model predicts the largest increase in galaxy overdensity, which increases with increasing UV luminosity limit, from $\sim3$ for $M_{\rm UV} < -18$ to $\sim 5$ for $M_{\rm UV} < -22$. 
In contrast, the high stochasticity model shows a more modest increase in the number of neighboring galaxies relative to mean density, from $\sim1.5-2$, for $M_{\rm UV} < -18$ to $-22$. 
The larger increase in the number of fainter galaxies around a bright galaxy for the high luminosity model again is a result of brighter galaxies residing in more massive halos compared to the high stochasticity model, and more massive halos being more biased tracers of the density field \citep{Bond1991}.
The amplitude of the increase in the number of neighboring galaxies is consistent with the change in the expected density distribution shown in the left panel of Fig.~\ref{fig:full_theory_plot}.

\subsection{Clustering prediction of individual fainter neighbors}
\label{sec:individual_analytic}

As we demonstrate directly from simulations in Sec.~\ref{sec:radial_cumulative}, the number of neighboring galaxies alone is too broadly distributed -- due to Poisson and cosmic variance -- to meaningfully distinguish between the two models. In analogy to the angular correlation function probing the clustering of galaxies for ensembles of galaxies, we can estimate the clustering of the neighbors around individual bright galaxies. We do this by extending the formalism presented in Section~\ref{sec:analytic} to predict the distribution of expected distances from the bright galaxy to its neighbors of a given UV magnitude.

The key physical link enabling the prediction of distances to galaxy neighbors is that the clustering strength --- and hence the separation between the bright galaxy and its neighbors --- is set by the effective bias of the population of such galaxies (both bright and faint) relative to the underlying matter field, which is itself a function of their halo mass. Let the population of the brightest galaxies in a given field be denoted by $B$, and the fainter population by $F$. For each population, we derive the effective bias:

\begin{equation}
b_i = \frac{\int_{M_{\rm UV,i}} \textrm{d}M_{\rm UV} \int_{M_{\rm min}}^{\infty} \textrm{d}M_h b(M_h) \cdot \frac{{\rm d}n}{{\rm d}M_{\rm UV}} \cdot p(M_{\rm h}\,|\,M_{\rm UV})}{\int_{M_{\rm UV,i}} \textrm{d}M_{\rm UV} \int_{M_{\rm min}}^{\infty} \textrm{d}M_h
\frac{{\rm d}n}{{\rm d}M_{\rm UV}} \cdot  p(M_{\rm h}\,|\,M_{\rm UV})}
\end{equation}

where $i\in\{B,F\}$, and $b(M_{\rm h})$ is the linear bias from \citet{Tinker2010}. The effective galaxy bias quantifies how strongly galaxies, as a population, trace the underlying dark matter distribution. We can then write the linear cross-correlation between the two populations as:

\begin{equation}
    \xi_{B,F} \approx b_B b_F \xi_{mm},
    \label{eq:xi_deriv}
\end{equation}

where $\xi_{mm}$ is the linear matter auto-correlation function, which can be readily obtained from the matter power spectrum.

From Eq.~\ref{eq:xi_deriv}, we can obtain the density of faint galaxies within a radial distance $r$ around bright galaxies as:

\begin{equation}
    \rho_F(r) = 4 \pi r^2 \cdot \overline{n}_F  \cdot \left(1 + \xi_{B,F}(r)\right),
\end{equation}

where $\overline{n}_F$ is the average number density of galaxies of population $F$. This enables us to estimate the expected number of galaxies of population F within some radius $R$ around a bright galaxy:

\begin{equation}
    N(<R) = \int_0^R \rho_F(r)\, \textrm{d}r.
\end{equation}

From this, we can compute the probability distribution of separations between bright galaxies and their fainter neighbors.
We first compute the survival function, which gives the probability that no galaxy from population $F$ lies within a radius $R$, under the assumption of a Poisson distribution of galaxy counts, $\exp[-N(<R)]$. This approximation is valid on scales larger than the virial radii ($R>0.05$--$0.25$\,cMpc for the halo mass range considered, $M_{\rm h}\approx[10^{8.5},10^{11.5}]M_{\odot}$), where the positions of individual halos are determined primarily by the large-scale density field rather than halo exclusion. The cumulative distribution function of the nearest-neighbor distance is then $1 - \exp[-N(<R)]$, and the corresponding probability distribution function of nearest-neighbor distances is obtained by differentiating this expression with respect to $R$:

\begin{equation}
    \mathcal{P}_F(R\,|\,B) = \frac{\textrm{d}}{\textrm{d}R}\left(1- e^{-N(<R)}\right)
\end{equation}

In the right panel of Fig.~\ref{fig:full_theory_plot}, we show this distribution of distances to the nearest galaxy, from a bright galaxy with $M_{\rm UV,B}=-21.5$, assuming the faint neighbors are brighter than $M_{\rm UV,F}<-18.5$. We clearly see the high luminosity model predicts shorter distances to the nearest neighbor, compared to the high stochasticity model. This reflects the larger effective bias of both the bright and neighboring galaxy populations: at fixed UV luminosity, they occupy more massive halos, which have a higher effective bias and therefore cluster more strongly. 
As a consequence, we are more likely to find a neighbor closer to the bright galaxy in the high luminosity model than in the high stochasticity model. 
Conversely, finding no nearby neighbors, with the nearest neighbor located at a large separation ($R\gtrsim4$\,cMpc), suggests that the bright galaxy resides in a lower-mass halo and therefore favors the high stochasticity model.
This demonstrates that individual distances encode the clustering signal which can isolate the information about the dark matter halos of the galaxies.

The analytic formalism presented here can be extended to construct analytic likelihoods for the number counts of fainter neighbors and their radial distance. 
However, there are several limitations of this analytic approach. 
In particular, it does not easily account for the large Poisson variance due to the low number of galaxies expected at $z\gtrsim10$ (see also Sec.~\ref{sec:radial_cumulative}), nor cosmic variance arising from fluctuations in the underlying matter density on both smaller and larger scales than the radius of the field, both of which impact the expected number and spatial separations of neighboring galaxies. 
Properly accounting for both sources of uncertainty requires forward-modeling realistic observations, allowing the total variance to be captured without introducing systematic biases.
Another limitation is that this approach assumes 3D radial distances are known for all galaxies, which requires spectroscopic redshifts. 
However, due to the expense of spectroscopy of faint galaxies at $z>10$, so far mostly photometric information is available \citep{Tacchella2023, Whitler2025, Hainline2026, McLeod2026}. In addition, the formalism assumes idealized spherical volumes and cannot account for the complex footprints of real observations.

These limitations motivate a more realistic 3D modeling approach: turning to simulations to forward-model JWST observations mimicking the environments of high-redshift galaxies such as GN-z11 and GS-z14-0.

\section{Empirical methodology}
\label{sec:methodology}

Generating realistic 3D galaxy locations allows us to forward-model JWST observations and understand the clustering at $z\gtrsim 10$, overcoming limitations of our analytic model.

In this work, we employ 21cmFAST \citep{Mesinger2011, Murray2019},  which  generates density fields using second-order Lagrangian perturbation theory \citep{Scoccimarro1998}. Dark matter halos are then generated by sampling the conditional mass function \citep{Sheth2001} on a cell-by-cell basis \citep[21cmFASTv4,][]{Davies2025}, producing halo catalogs, for the halo masses ($\gtrsim10^{9}M_{\odot}$) and spatial scales ($\gtrsim1$\,cMpc) of interest in this work, with a mass function and correlation statistics consistent with N-body simulations \citep{Sheth1999, Qiu2021}. Furthermore, an entire simulation (with a size of $500$ cMpc and a resolution of $1$ cMpc) can be performed in hours down to redshift $z=8$. This short computation time allows us to probe large volumes to generate a statistically significant number of massive halos hosting the brightest galaxies ($>100$ $M_{\rm UV} \lesssim -21.5$ galaxies at $z=10.5$) with precise locations of all halos down to $10^8\,M_{\odot}$ on scales larger than $1$ cMpc. Achieving a similar combination of large volume and mass resolution with dark-matter-only N-body simulations would be computationally prohibitive \citep{Springel2005, Klypin2011}. The large volume and low halo mass threshold in our 21cmFAST simulations allow us to explore different astrophysical and cosmological parameters, compare to the other large-scale observables, and give us a better handle on the cosmic variance (through repeated simulations with different random seeds).

\begin{figure*}
    \centering
    \includegraphics[width=1.0\linewidth]{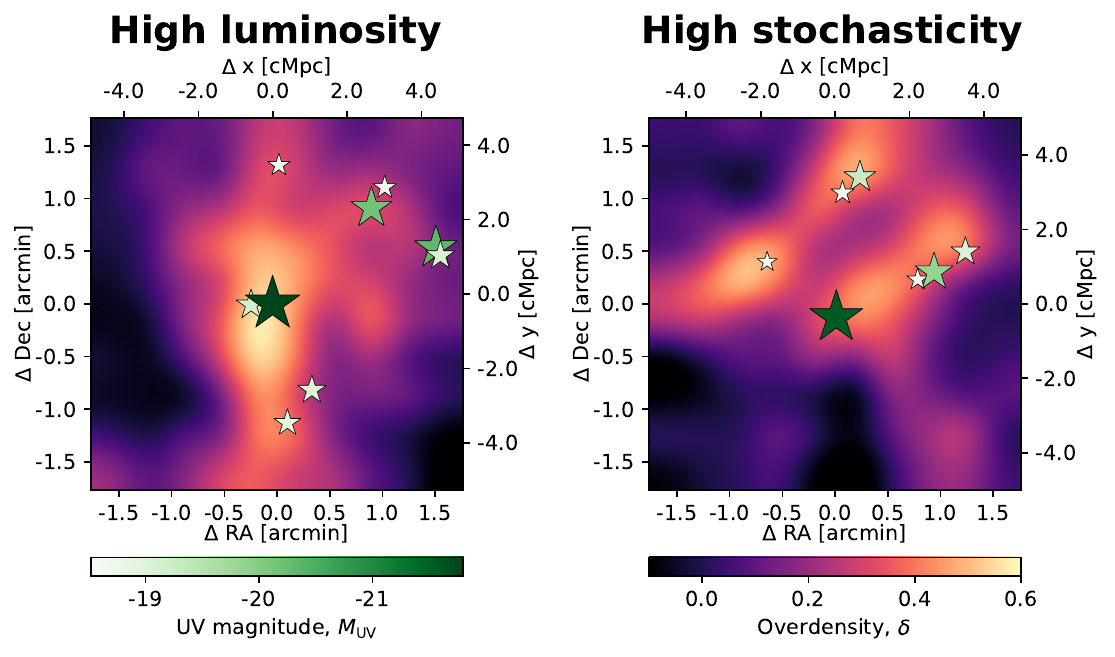}
    \caption{Examples of forward-models of environments of bright galaxies (with side-length of $10$\,cMpc, corresponding to $\sim3.6$\,arcmin, and line-of-sight (LoS) length of $10$\,cMpc, corresponding to the redshift interval of $\Delta z\approx0.05$) with neighboring galaxies overlaid on the overdensity field (averaged over a $1$\,cMpc scale) centered on the brightest galaxy ($M_{\rm UV,0} \approx -21.5$). Star size and color both scale with galaxy UV luminosity ($-M_{\rm UV}$, colorbar below left panel).
    The left panel shows an example generated using the high luminosity model, and the right panel shows an example generated using the high stochasticity model. These examples demonstrate the different environments around the brightest galaxy in the two models.}
    \label{fig:cutout_examples}
\end{figure*}

We obtain UV magnitudes of galaxies by sampling the conditional distribution $p(M_{\rm UV}\,|\,M_h)$ described in Sec.~\ref{sec:analytical_framework} for each halo in the simulations down to the mass of atomic cooling halos, $\log(M_{\rm h}/M_{\odot})>8.5$), which are expected to be forming stars \citep{Barkana2001, Sobacchi2013}\footnote{We assume one galaxy per dark matter halo. This assumption ignores the one-halo clustering contribution from subhalos. This could affect the nearest-neighbor distribution on very small scales ($\lesssim 2$\,arcsec, or equivalently $\approx0.1$\,cMpc). We therefore repeat the analysis in this section excluding the innermost $0.1$\,cMpc around each bright galaxy. Fewer than $1$\% of systems in the high luminosity model have a neighbor within this distance, and an even smaller fraction in the stochasticity model, leaving our conclusions unchanged.}. For the two models in Section~\ref{sec:analytical_framework}, we generate galaxy catalogs at $z\in\{8,10.5,12.0,14.0\}$. To robustly sample the stochastic mapping between halo mass and UV luminosity, we repeat this procedure 200 times by independently drawing $M_{\rm UV}$ realizations for every halo, thereby generating 200 Monte Carlo realizations of each catalog. This results in a sample of $\sim30,000$ UV-bright galaxies with $M_{\rm UV}<-21.5$ at $z=10.5$, and their environments.

We process the galaxy catalogs as follows. We start by identifying all UV-bright galaxies in the catalog at a redshift $z$ brighter than a given $M_{\rm UV,0}$ (see Section~\ref{sec:analytical_framework}). We define a cutout around each bright galaxy using a rectangular volume (i.e. corresponding to a square on the sky) defined by its extent on the sky, $L$, and along the line of sight, $L_z$. Within this volume, we consider all galaxies brighter than a survey limiting magnitude, $M_{\rm UV, lim}$. In summary, a cutout is defined by the five quantities:

\begin{itemize}
    \item $z \rightarrow$ central redshift of the cutout
    \item $M_{\rm UV, 0} \rightarrow$ UV magnitude of the central bright galaxy used as a condition for the environment 
    \item $M_{\rm UV, lim} \rightarrow$ limiting UV magnitude below which we observe \textit{all} of the galaxies, i.e. assuming $100\%$ completeness brighter than that UV magnitude
    \item $L \rightarrow$ the length of the field on the sky (assuming a square pointing)
    \item $L_z$ or equivalently $\Delta_z \rightarrow$ length of the field along the line-of-sight
\end{itemize}

We motivate the fiducial values of the cutouts using observations of GN-z11 \citep{Oesch2016, Bunker2023, Tacchella2023}. GN-z11 is the brightest known galaxy at $z>10$, with $9$ photometric candidates in its vicinity \citep{Tacchella2023}. Our fiducial bright sample thus is set at $z=10.6$, with $M_{\rm UV,0} = -21.5$, and a limiting magnitude of the faint sample, $M_{\rm UV,lim} \leq -18.5$, approximately corresponding to the UV magnitude of the faintest candidate in \citet{Tacchella2023} (for which completeness is close to $100\%$, see also \citet{Whitler2025}). We assume $L=10$ cMpc, matching the field of view described by \citet{Tacchella2023}. For intuition, we start by assuming $L_z = L$ ($\Delta z =0.05)$, i.e., cubical cutouts, before building realism by considering photometric redshifts which typically span a larger range in $L_z$ in Section~\ref{sec:ang}. Finally, we construct the cutouts for both UV luminosity-halo mass models from Section~\ref{sec:analytical_framework}.

We show an example of the cutouts in  Fig.~\ref{fig:cutout_examples} for both the high luminosity and high stochasticity models. The cutouts are centered on the brightest galaxy in the cutout (with $M_{\rm UV,0}\approx-21.5$ for both models), with galaxies colored according to their UV magnitude. In the background we also show the underlying dark matter density averaged over $1$\,cMpc scales. 

We find galaxies trace the underlying matter density in both models. 
However, bright galaxies are clearly hosted in different environments in the two models. In the high-luminosity model, the bright galaxy and most of its neighbors reside within a single prominent overdense region, tracing the same large-scale structure. In contrast, in the high-stochasticity model the bright galaxy can lie in a comparatively moderate overdensity, while some of the nearby bright galaxies originate from distinct overdense regions within the same field. In this case, apparent proximity does not necessarily imply a shared host environment.

From the cutouts presented in this subsection we can then extract summary statistics to probe clustering. We explore several of the summaries in the following subsections.

\begin{figure}
    \centering
    \includegraphics[width=1.0\linewidth]{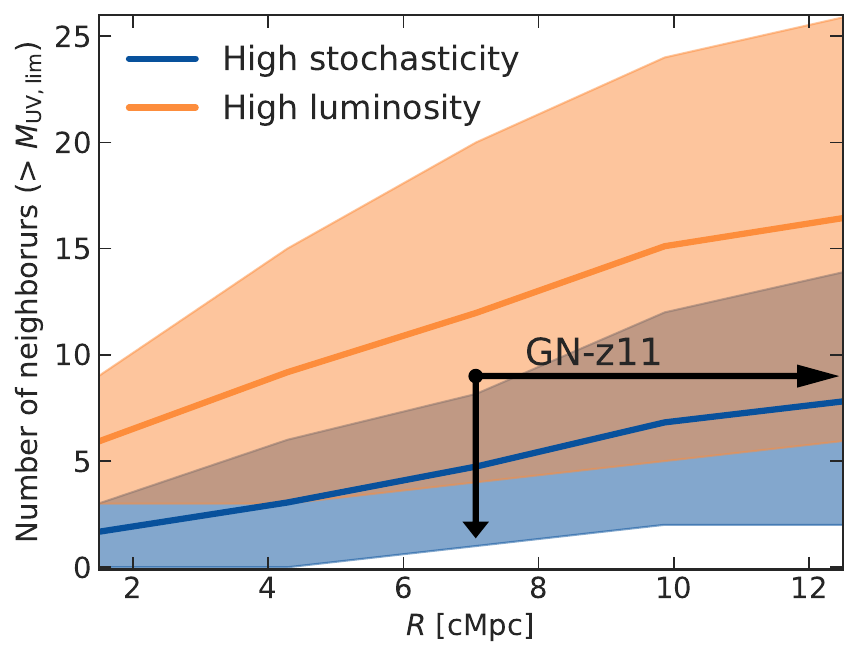}
    \caption{Number density of neighboring galaxies as a function of distance for the high luminosity (orange) and high stochasticity (blue) models. Solid lines show the median across all cutouts, while shaded regions indicate the $68\%$ confidence intervals. The observed GN-z11 photometric neighbors are also shown, with distances treated as lower limits in terms of radius due to the unknown line-of-sight positions of the galaxies and upper limits in terms of total number of neighbors. Although the models predict different neighbor densities, the large cosmic-variance uncertainty limits the ability of this statistic to distinguish between them in the case of GN-z11.}
    \label{fig:num_dens_sim}
\end{figure}

\subsection{Radial distribution of neighbors}
\label{sec:radial_cumulative}
Similarly to Section~\ref{sec:analytic}, we first forecast the number of galaxies in a given volume, which is the simplest way to quantify the overdensity of a region \citep{Steidel1998, Hogg2003, Kauffmann2004, Ren2018, Helton2024, Kashino2023, Wu2026}. However, Poisson fluctuations in the number counts of galaxies and cosmic variance due to unknown underlying dark matter density lead to a large total variance in the expected counts, which must be accounted for.

To estimate the total variance in galaxy number counts, we compute the cumulative number of galaxies, $N$, above $M_{\rm UV, lim}$ around a bright galaxy with $M_{\rm UV,0}$ within a 3D radius $R$. We repeat the procedure for the two models, across all cutouts ($\sim30,000$ per model) created using the procedure outlined above. The mean radial distribution of neighboring galaxies is shown in Fig.~\ref{fig:num_dens_sim}, along with the $68\%$ confidence interval (C.I.).

Fig.~\ref{fig:num_dens_sim} demonstrates the high luminosity model predicts a larger mean number of neighboring galaxies at all distances compared to the high stochasticity model (even though the average over the whole simulation is the same for both models). The mean relation is larger by a factor of $\sim2$ for the high luminosity model. This increase matches well with the prediction from Fig.~\ref{fig:full_theory_plot} on scales of $R\sim10$ cMpc.

However, the distribution of neighbor counts within a given separation is broad, making this statistic challenging for distinguishing between galaxy formation models. The width of the distribution arises from the Poisson fluctuations in the number counts, as well as the diversity of environments around the bright galaxies. 
This issue was previously discussed by \citet{Ren2018}, who proposed using the number of neighbors of bright galaxies at $z\sim8.5$ to constrain the level of UV luminosity-halo mass scatter, but found wide distributions of the number of neighbors around $m_{AB} \sim 25$ galaxies. 
\citet{Munoz2008} reached a similar conclusion in the context of using the number of neighbors as a way to confirm the redshift of a massive galaxy at $z=6.5$, also finding a wide distribution of the number of neighbors. 
At $z>10$ the situation is even more challenging than at $z\sim8.5$ due to the lower number of both bright galaxies and their photometric neighbors with increasing redshift. 
Indeed, we note that the number of neighbors around GN-z11 places it in this ambiguous regime. \citet{Tacchella2023} reported nine photometric candidates around GN-z11 (within a radial distance of $\sim7.5$ cMpc). 
Naively placing these nine candidate neighbors in Fig.~\ref{fig:num_dens_sim} (assuming all are physically associated with GN-z11) we find the observed number of neighbors is within the $68\%$ confidence interval of both models (black arrow). Importantly, the number density is treated as an upper limit as the radial distances of the neighbors are not known (which may increase the total separation from the bright galaxy), and some of them might be interlopers (decreasing the total number of true neighbors).
This implies the radial distribution of GN-z11's neighbors is not sufficiently informative to distinguish between the two extreme models. While measuring an extremely high number of photometric candidates ($N\gtrsim20$), or similarly zero neighbors, would provide clearer evidence to discriminate between the models, current observations do not find such extreme values, motivating statistics beyond simple number counts. 

The inability of neighbor counts alone to distinguish between the two models is a manifestation of the same degeneracy that affects the UV luminosity function. Both high luminosity and high stochasticity models can reproduce similar abundances of both bright galaxies and fainter photometric candidates on scales corresponding to JWST photometric surveys \citep{Donnan2024, Whitler2025}. On the spatial scales considered here, this degeneracy is also present, with elevated uncertainties dominated by Poisson fluctuations and cosmic variance, which broaden the distribution of neighbor counts. As a result, two physically distinct models can predict very similar neighbor counts. The key additional information lies not in how many neighbors are present, but in how they are distributed spatially around a bright galaxy, similar to the angular correlation function which describes this for an ensemble of galaxies. Motivated by this, we therefore turn to statistics which include more spatial information.

\subsection{Distances between closest galaxies}

We now consider alternative statistics that more directly probe the spatial distribution of galaxies. Information about galaxy clustering is encoded in the separations between individual galaxies. The traditional estimate of clustering, angular correlation functions, extracts this information by averaging over a large number of galaxy pairs. Here, however, we show that individual galaxy pairs can already provide meaningful constraints. The right panel of Fig.~\ref{fig:full_theory_plot} shows that two galaxy formation models from Sec.~\ref{sec:analytic} predict different individual distances from the bright galaxy.
A similar approach was proposed by \citet{Banerjee2021}, who used distribution functions of the distances to galaxies' nearest neighbors to infer cosmological parameters.
A complementary approach was recently used by \citet{Wu2026b}, who measured relative velocities of spectroscopically confirmed close pairs at $z\sim7$ to derive dynamical halo mass estimates.
Here we demonstrate that the distance to a bright galaxy's nearest neighbor can be a powerful summary statistic for distinguishing our galaxy formation models.

Similarly to the procedure described in Section~\ref{sec:individual_analytic}, we select nearest neighbors of bright galaxies from the galaxy catalogs created in Section~\ref{sec:methodology}.
For each cutout around a bright galaxy, we measure the distance $d_1$ to the closest neighbor around the bright galaxy. More generally, the distance to the $i$-th closest neighbor from a bright galaxy is denoted $d_i$.

\begin{figure}
    \centering
    \includegraphics[width=1.0\linewidth]{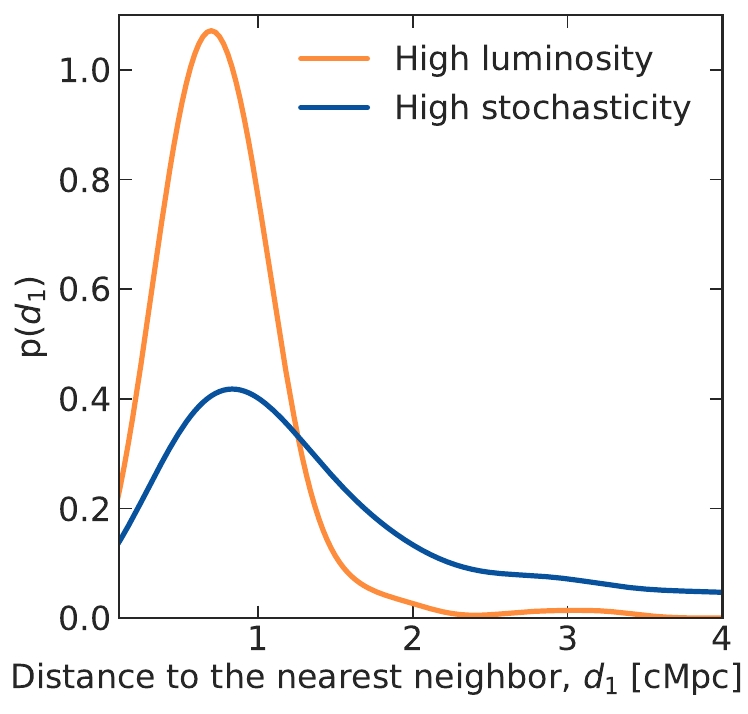}
    \caption{Probability distribution of the distance to the closest neighbor $d_1$ in $10 \times 10 \times 10$ cMpc$^3$ cutouts for the high luminosity (orange) and high stochasticity (blue) models. The high luminosity model shows a distribution peaked at small separations, while the high stochasticity model exhibits larger typical separations and an extended tail toward higher values of $d_1$.}
    \label{fig:PDF_global}
\end{figure}

In Fig.~\ref{fig:PDF_global} we present the probability distributions of the nearest-neighbor separations, $p(d_1)$, for all cutouts satisfying the selection criteria (namely, that they are brighter than $M_{\rm UV,lim} = -18.5$, and are within the cutout defined by the side length $L =10 $\,cMpc and LoS length $L_z=10$\,cMpc), for the two models. 
The high luminosity model predicts a high probability of the nearest neighbor having small separation $<1\,\mathrm{cMpc}$ from the bright galaxies, and declines rapidly beyond $1.5$ cMpc. In contrast, the nearest-neighbor separation distribution in the high stochasticity model peaks at larger separations, has a wider distribution around the peak, and displays an extended tail to $d_1 > 3\,\mathrm{cMpc}$. Quantitatively, the probability of finding a neighbor at $d_1>3$ cMpc decreases from $\approx 35\%$ in the high stochasticity model to $\approx 3\%$ in the high luminosity model. This difference provides a direct means of distinguishing between the models: a bright galaxy with a distant nearest neighbor ($d_1>3$\,cMpc) would favor the high-stochasticity scenario, while a close neighbor ($d_1<1.5$ cMpc) would instead favor the high-luminosity model.

Additionally, the number of neighbors and their separations can be used jointly, as both encode information about the environment of a bright galaxy. In Appendix~\ref{sec:cond_on_N}, we condition the nearest-neighbor separation, $d_1$, on the number of fainter neighbors, $N$, and show that it modifies the expected probability distribution of $d_1$ in the following way. Namely, in the high luminosity model, the distribution of $d_1$ does not change significantly based on the number of photometric neighbors, as bright galaxies in that model are always more likely to be in overdensities. Meanwhile, in the high stochasticity model, the distribution of $d_1$ becomes wider for bright sources with a lower number of neighbors (see Fig.~\ref{fig:CDF_by_N}) -- indicating a less massive halo in a less pronounced overdensity. The joint information from $d_1$ and total number of neighbors thus has the potential to increase the belief in either of the two models.

\subsection{How observational choices shape the clustering signal} \label{sec:obs_choice}

As the clustering of halos depends on their mass and redshift \citep{Bond1991, Tinker2010}, the constraining power of the nearest-neighbor distance, $d_1$, depends on both the redshift and luminosity range of our observations.
In particular, both the UV magnitude of the bright galaxy $M_{\rm UV,0}$, and the limiting UV magnitude of the neighbors $M_{\rm UV,lim}$ will alter the clustering signal.
In this subsection, we explore how these parameters shape the $d_1$ distributions and identify observational regimes where the models can be most easily distinguished.

In the following, we use the simulations described in Sec~\ref{sec:methodology}, and vary the UV magnitude of the bright galaxy, $M_{\rm UV,0}$, which determines the cutout selection (Sec.~\ref{sec:muv0_change}), and UV magnitude of the neighboring galaxies, $M_{\rm UV,lim}$, which determines the depth for each cutout (Sec.~\ref{sec:muvlim_change}). We also vary the redshift of the simulation and reconfigure the models to match the JWST-observed UVLF in Sec.~\ref{sec:red_dep}.

\subsubsection{Dependence on the luminosity of the brightest galaxy, $M_{\rm UV,0}$}
\label{sec:muv0_change}

\begin{figure*}
    \centering
    \includegraphics[width=1.0\linewidth]{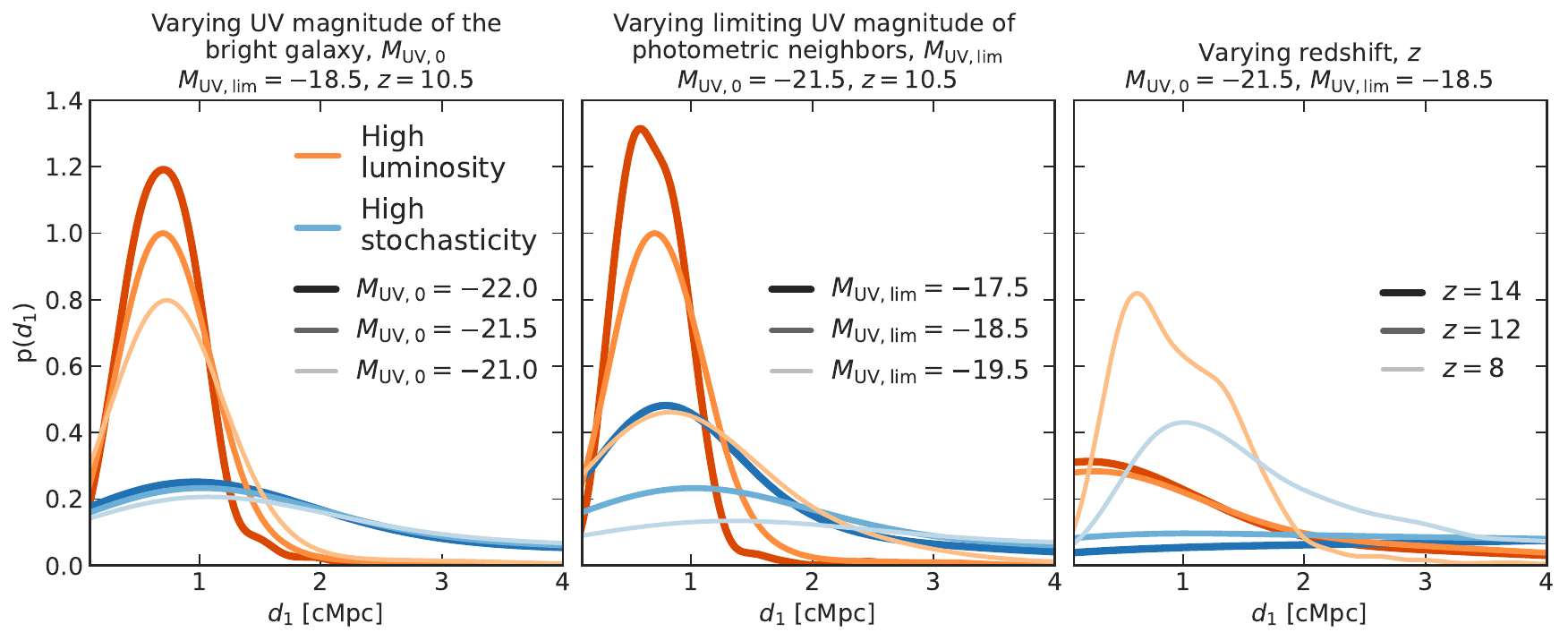}
    \caption{Effect of observational parameters on the distance to the closest neighbor for the high luminosity (orange) and high stochasticity (blue) models. \textit{Left panel:} Varying the UV magnitude of the central galaxy, $M_{\rm UV,0}$. Brighter central galaxies probe more strongly clustered regions, enhancing the differences between the models. \textit{Middle panel:} Varying the limiting UV magnitude of the neighbor population, $M_{\rm UV,lim}$. Including fainter galaxies decreases the distance to the closest neighbor for both models, reducing their distinguishability. Conversely, selecting only brighter galaxies shifts neighbors to larger distances. \textit{Right panel:} Varying redshift. At $z=8$, both models predict short distances to the closest neighbor. At higher redshifts, the nearest neighbor is less likely to be physically associated with the same overdensity, leading to flatter distributions.}
    \label{fig:muvlim_change}
\end{figure*}

We now discuss how the nearest-neighbor separation depends on the luminosity of the central bright galaxy, $M_{\rm UV,0}$. 
The left panel of Fig.~\ref{fig:muvlim_change} shows the distribution of the nearest neighbor separations obtained by selecting bright galaxies in our simulations from a range of UV magnitudes: $M_{\rm UV,0} \in \{-21.0, -21.5, -22.0\}$ (corresponding to apparent magnitudes $m_{AB}\approx 26.6, 26.1, 25.6$ at $z=10.5$). In all three cases we fix the survey limiting magnitude, $M_{\rm UV,lim} = -18.5$ ($m_{AB}\approx 29.1$), and redshift $z=10.5$.

The distributions of distances to the nearest neighbors do not strongly depend on the UV magnitude of the central bright galaxy, $M_{\rm UV,0}$, but generally become easier to separate as we consider brighter central galaxies. In the high luminosity model, the distribution peaks at $d_1 \approx1$\,cMpc for all $M_{\rm UV,0}$, with a slight shift towards lower values of $d_1$ and a narrowing of the distribution for brighter $M_{\rm UV,0}$. The high stochasticity model similarly shows a slight narrowing of the peak of the distribution for increasingly luminous $M_{\rm UV,0}$, but for the magnitude range we test here all the distributions still have a significant tail above $d_1>3$\,cMpc. 
This narrowing of the peak of $p(d_1)$ in both models arises because, as we select brighter central galaxies, we are, on average, probing more massive halos (see Fig.~\ref{fig:pmuv}), which trace more clustered environments. 
As we will show in Section~\ref{sec:ks_test}, this effect becomes more pronounced as we go to higher redshifts.
Nevertheless, because the $M_{\rm UV}$--$M_{\rm h}$ relation is so broad in the high stochasticity model, the brightest galaxies in this model are still hosted in a broad range of halos so the $d_1$ distribution does not change dramatically. 
The fact that the distribution of $d_1$ does not significantly depend on the central galaxy luminosity for $M_{\rm UV,0} \leq -21$, in either model, means that galaxies across this luminosity range can be treated as a single population, enabling us to build a larger sample for our analysis, as we explore in Section~\ref{sec:ks_test}.

\subsubsection{Dependence on survey magnitude limit, $M_{\rm UV,lim}$}
\label{sec:muvlim_change}
The limiting UV magnitude of the survey, $M_{\rm UV,lim}$, sets the number density of neighbors as well as their bias with respect to the bright galaxy. Thus, the limiting magnitude affects the probability that a neighbor is associated with the same overdensity as the bright galaxy: if the magnitude limit is too high, we may not be probing galaxies in the same overdensity.

In the middle panel of Fig.~\ref{fig:muvlim_change} we show how varying the limiting magnitude, $M_{\rm UV,lim}$, impacts the $d_1$ distribution, fixing the bright galaxy magnitude $M_{\rm UV,0}=-21.5$ ($m_{AB}\approx 26.1$ at $z=10.5$). As the survey limit deepens to include progressively fainter galaxies, the distribution of $d_1$ shifts to smaller separations,
as the higher number density of neighbors increases the probability that a neighbor traces the same overdensity as the bright galaxy. 
For $M_{\rm UV,lim} = -17.5$ ($m_{AB}\approx 30.1$), almost all bright galaxies contain a neighbor within the resolution element ($1$\,cMpc). A single observation of a neighbor within $d_1<2$\,cMpc would lead to only a low preference for the high luminosity model.
On the other hand, raising the magnitude limit reduces the number density of neighbors,
increasing the probability that the closest neighbor is associated with a physically separate overdensity from the bright galaxy (see Fig.~\ref{fig:cutout_examples}). 
As a result, for the brightest limiting magnitude we consider, $M_{\rm UV,lim} = -19.5$ ($m_{AB}\approx 28.1$), the distributions for both models exhibit long tails towards large distances. However, an observation of a neighbor at short distance, $d_1<2$\,cMpc, would still favor the high luminosity model.

Overall, this implies that for a given UV magnitude of the brightest galaxy, $M_{\rm UV,lim}$, there will be an optimum survey limit for which the two models' nearest-neighbor separation distributions can be most easily distinguished. We will discuss this in more detail in Sec.~\ref{sec:ks_test}.

Throughout this analysis, we assume 100\% completeness down to the limiting magnitude $M_{\rm UV,lim}$, i.e. that all galaxies down to that magnitude are detected.
In practice, follow-up observations are necessary to confirm the photometric candidates and to assess contamination by interlopers.
We assessed the impact of incompleteness by randomly removing 10\% of the neighbor galaxies in each cutout and recomputing the $d_1$ distributions, where the completeness level is consistent with \citet{Whitler2025}. We found this had a negligible impact on our results at all redshifts considered.

\subsubsection{Redshift dependence}
\label{sec:red_dep}

The relation between galaxy UV luminosity and halo mass evolves with redshift as the growth of galaxies is shaped by the evolving supply of gas and the resulting star-formation histories \citep{Behroozi2013, Speagle2014, Wechsler2018}. At higher redshifts, increasing accretion rates lead to higher star-formation rates at fixed halo mass, causing galaxy of a given UV luminosity to reside in progressively lower-mass halos \citep{Mason2015, Sun2023, Feldmann2025}. Accurately predicting the clustering of bright galaxies across redshifts therefore requires accounting for the evolution of the UV magnitude--halo mass relation, including both the evolution of its median relation and the scatter around it.

We account for the evolution of the UV magnitude - halo mass relation by adopting the relations presented in  \citet{Mason2015, Mason2023} as the median of the probability distribution (i.e., $\mu_\mathrm{UV}(M_{\rm h}, z)$) for each redshift (i.e., $z\in\{8,12,14\}$). These relations capture how rising accretion rates result in galaxies of fixed UV magnitude residing in increasingly low mass halos at higher redshifts. 
For each model, we assume these median $\mu_\mathrm{UV}(M_{\rm h}, z)$ relations, and adopt the stochasticity described in Sec.~\ref{sec:analytical_framework}. 
We calibrate both models to reproduce the same UVLF at each target redshift. This ensures that any differences in the predicted clustering arise solely from differences in the galaxy-halo connection, rather than from variations in the abundance of galaxies.

Thus, for the high luminosity model, we assume a constant level of stochasticity with redshift (i.e. a mass-independent $\sigma_{\rm UV}=0.3$\,mag), and adjust the shift, $\mu_{\rm UV, shift}$, of the median UV luminosity-halo mass relation, to match the UVLF at each redshift as necessary.
For the high stochasticity model, we hold the mass-dependence slope of $\sigma_{\rm UV}(M_h) = \sigma_a(\log(M_{\rm h}/M_{\odot})-12) + \sigma_b$ fixed at $\sigma_a=-0.34$, and adjust only its normalization $\sigma_b$ — the value of $\sigma_{\rm UV}$ at $\log(M_{\rm h}/M_{\odot})=12$ — to explore how increasing stochasticity at fixed halo mass with redshift impacts clustering. 
Additionally, as introducing stochasticity impacts the normalization of the UVLF \citep{Ren2019}, we also adjust the shift of the median UV luminosity-halo mass relation, i.e., $\mu_{\rm UV, shift}$, when required, for the stochastic model to match the observations and high luminosity LF model.

In the Appendix~\ref{sec:app_z_uvlf} we show the resulting UVLFs and the match with observations used to the calibrate the models. 
In summary, the adjustments for each model at each redshift are:

\begin{itemize}
    \item \textbf{$\mathbf{z=8}$}: 
    For the high luminosity model, the \citet{Mason2015,Mason2023} model already matches the observed UVLFs well \citep{Bouwens2021, Adams2024, Willot2024, Kreilgaard2026}, so we do not adjust the median of the $M_{\rm UV}$--$M_{\rm h}$ relation, i.e., $\mu_{\rm UV, shift}= 0.0$\,mag. 
    For the high stochasticity model, we change the normalization of the stochasticity, $\sigma_{b}$ to $0.6$\,mag, and the shift in the median to $\mu_{\rm UV, shift}= +0.8$\,mag, as required to match the observed UVLF (see above). 
    \item \textbf{$\mathbf{z=12}$}: 
    For the high luminosity model, the \citet{Mason2015,Mason2023} model underpredicts the UVLFs even more than at $z=10.5$ \citep{Willot2024, Donnan2024, Adams2024, Whitler2025}, so we thus further increase UV luminosity for a given halo mass, $\mu_{\rm UV, shift}=-1.2$\,mag.
    In the high stochasticity model, adopting the median $M_\mathrm{UV}(M_h, z=12)$ from \citet{Mason2023} with the same level of stochasticity as we used at $z=10.5$ (see Sec.~\ref{sec:analytical_framework}), fits the observed UVLF well, with no additional adjustments.
    \item \textbf{$\mathbf{z=14}$}: 
    For the high luminosity model, the \citet{Mason2015,Mason2023} model severely underpredicts the UVLFs \citep{Finkelstein2023, Donnan2024, Robertson2024, Whitler2025}, requiring us to increase UV luminosity by an order of magnitude for a given halo mass to match the observed UV LF ($\mu_{\rm UV, shift}=-2.5$\,mag).
    The high stochasticity model, which matched $z=10.5-12$, also underpredicts the $z=14$ UVLF, thus we increase the normalization of the stochasticity, $\sigma_{b}$ to $0.9$\,mag, to upscatter enough halos to match the observed UVLF, with no additional shift to the median (as at $z=12$).
\end{itemize}

The right panel of Fig.~\ref{fig:muvlim_change} shows the evolution of the nearest-neighbor distance distribution, $p(d_1)$, at $z=8$, $12$, and $14$ for fixed $M_{\rm UV,0}=-21.5$ and $M_{\rm UV,lim}=-18.5$. Similarly to the shape of the distribution function at $z=10.5$ (Fig.~\ref{fig:PDF_global}), the high stochasticity model (represented with shades of orange) shows a broader distribution of $d_1$ compared to the high luminosity model (shades of blue). 
We now describe each redshift in turn.
At $z=8$ (light blue and orange curves), the high luminosity model predicts that the nearest neighbor typically lies within $2$ cMpc, similar to the result at $z=10.5$. On the other hand, in the high stochasticity model, $p(d_1)$ is shifted to lower separations with respect to $z=10.5$ (Fig.~\ref{fig:PDF_global}), with a less pronounced tail to large separations.
This reflects hierarchical structure growth: at lower redshifts, galaxies of a fixed $M_{\rm UV,0}$ reside in more massive halos than at $z=10.5$, and therefore occupy more strongly clustered environments. As a result, the two model distributions become more similar, limiting the constraining power of a single summary statistic for this choice of $M_{\rm UV,0}$ and $M_{\rm UV,lim}$. 
However, as discussed in Section~\ref{sec:muvlim_change}, using a shallower survey limit for the neighbors can help distinguish the models, and the higher number density of bright galaxies at lower redshifts gives us a larger sample to estimate the probability distribution of $d_1$. In Section~\ref{sec:ks_test} we quantify the optimum survey depth and how many galaxies are needed to distinguish the two models.

At $z\ge12$, the two models show qualitatively different behavior. In the high luminosity model, the amplitude of the peak of the nearest-neighbor distribution decreases and an extended tail develops toward large distances ($d_1\sim6$--$8$\,cMpc). In contrast, the high stochasticity model predicts an almost flat $d_1$ distribution at $z=12$ and $14$.
The behavior of the high luminosity model results from the decreasing abundance of both bright galaxies and their neighboring galaxies at higher redshifts. Although galaxies of fixed UV luminosity occupy lower-mass halos at earlier times, the tight luminosity-halo mass relation still associates bright galaxies with relatively overdense regions. However, the reduced number density of potential neighbors lowers the probability of finding a physically associated companion, weakening the peak at small separations and producing a broader tail.
In the high stochasticity model, bright galaxies at $z=12$ and $14$ are increasingly likely to result from strong upward fluctuations in UV luminosity at fixed halo mass. These galaxies therefore occupy less overdense environments, reducing the excess probability of nearby photometric neighbors and causing the nearest-neighbor distribution to approach a nearly uniform distribution.
For this choice of $M_{\rm UV,0}$ and $M_{\rm UV,lim}$, the two models can therefore be distinguished: detecting a neighbor within $d_1<2$\,cMpc would favor the high luminosity model, while the absence of close neighbors would provide evidence for the high stochasticity model. Increasing the depth by lowering $M_{\rm UV,lim}$ (see middle panel of Fig.~\ref{fig:muvlim_change}) increases the probability of detecting associated neighbors and extends the regime where clustering can be used as a diagnostic. Identifying these optimal observational regimes is therefore essential, and we explore them in Subsections~\ref{sec:ks_test} and \ref{sec:future}.

\subsection{Angular separation of galaxies}
\label{sec:ang}

While we have shown that the radial separations between bright galaxies and their neighbors can be a powerful probe of clustering, most bright $z>10$ sources currently lack the deep spectroscopy of their neighbors required for accurate 3D separations.
Thus, we now extend our approach to photometric surveys, considering angular positions and photometric redshift uncertainties.

\begin{figure}
    \centering
    \includegraphics[width=1.0\linewidth]{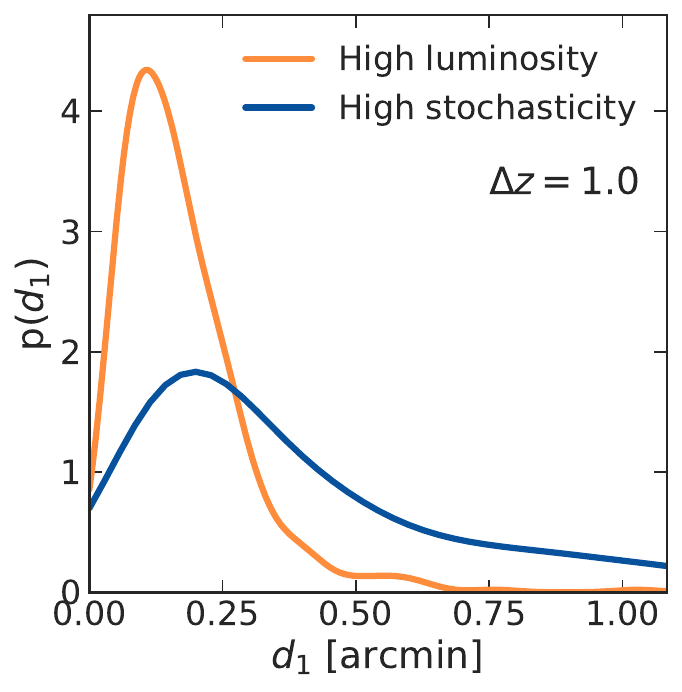}
    \caption{Probability distribution function of the angular distance to the closest neighbor for the high luminosity (orange) and high stochasticity (blue) models. As in the three-dimensional case (Fig.~\ref{fig:PDF_global}), the high luminosity model predicts systematically smaller separations, while the high stochasticity model exhibits a broader distribution with an extended tail toward larger distances.}
    \label{fig:ang_dist}
\end{figure}

To achieve this we now consider pencil beam volumes that are extended along the line of sight ($L_z$), and report angular separations rather than 3D distances.
As a fiducial value, we choose $L_z$ corresponding to a redshift width of $\Delta_z = 1.0$ (or $\approx 200$\,cMpc). This value is motivated by the typical redshift uncertainty associated with medium-band photometry \citep[see e.g.,][]{Suess2024, Adams2025, Eisenstein2026}. 

In Fig.~\ref{fig:ang_dist} we show the distribution of angular distances to the closest neighbor for the two models, assuming the UV magnitude of the bright population $M_{\rm UV,0}=-21.5$, and of the photometric neighbors $M_{\rm UV,lim}=-18.5$, at $z=10.5$. 
Encouragingly, we recover the same trends as we found in the three-dimensional case (see Fig.~\ref{fig:PDF_global}): the high luminosity model predicts shorter distances to the nearest neighbor than the high stochasticity model. 
This demonstrates that projection effects resulting from the redshift uncertainties do not significantly alter the differences between the two distributions. 
This can be understood from the small expected number of galaxies that are close in projected distance: integrating the UVLF over the cylinder of radius $R=1.5$\,cMpc and $L_z = 200$\,cMpc around the bright galaxy we expect only $\sim0.3$ additional galaxies. 
Since most of the probability density of the nearest-neighbor distance is contained within this radius (see Fig.~\ref{fig:PDF_global}), additional projected galaxies included in the pencil beam volume have little effect on the resulting distribution.

The distributions shown in Fig.~\ref{fig:ang_dist} motivate a survey strategy for measuring the clustering of photometric neighbors around bright galaxies. For both the high luminosity and high stochasticity models, the majority of the nearest-neighbor probability is contained within $R\leq1.1$ arcmin of the bright galaxy. Therefore, a $2.2\times2.2$ arcmin$^2$ region, corresponding to a single NIRCam module, is sufficient to characterize the local environment. The question then only is how deep the observations have to go, which determines the faintest detectable neighbors, and how many bright galaxies need to be targeted, which we explore in the next section.

\subsection{Optimizing the observational strategy}
\label{sec:ks_test}

\begin{figure*}
    \centering
    \includegraphics[width=\linewidth]{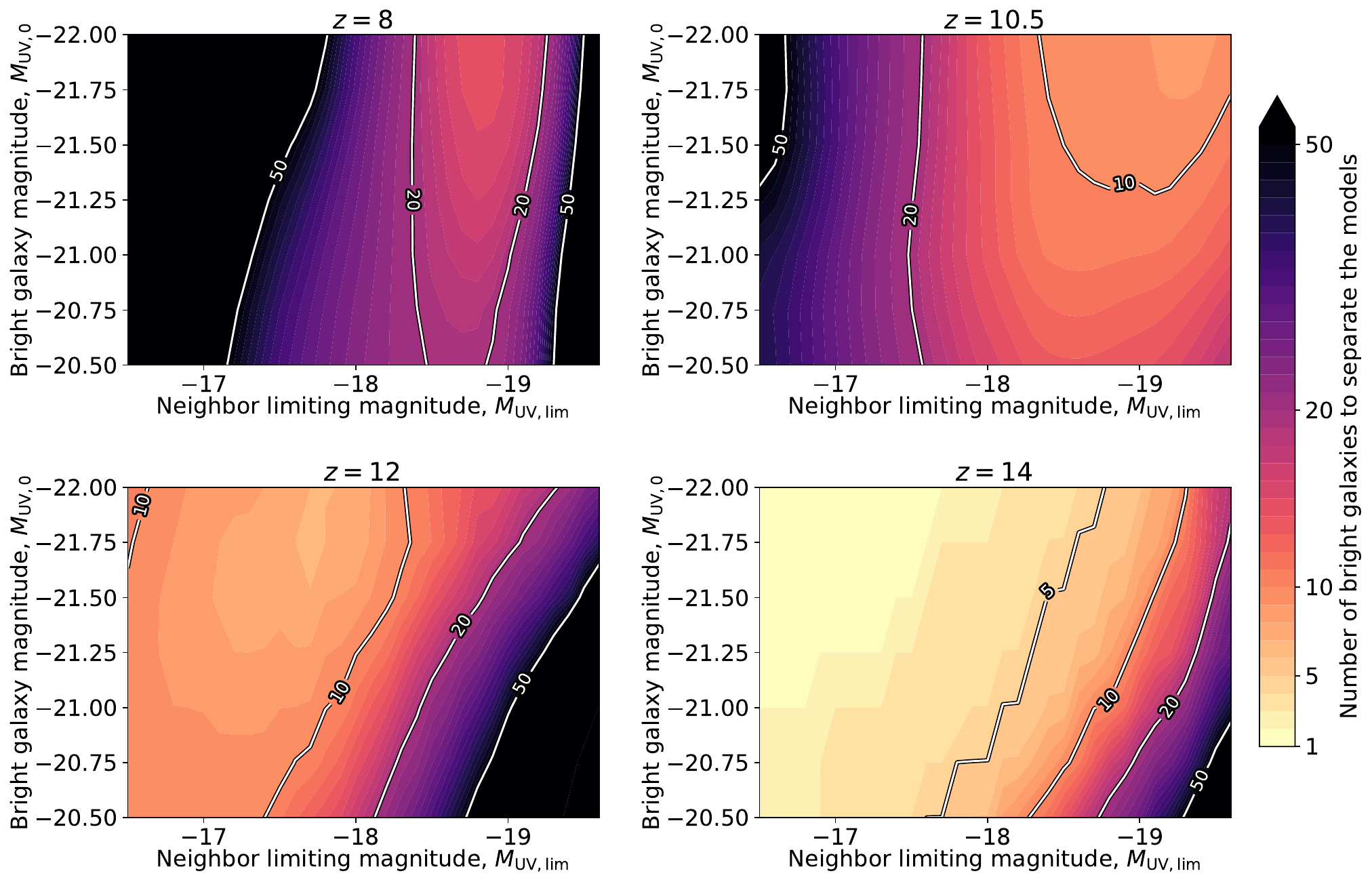}
    \caption{Number of bright-galaxy pointings required to distinguish between the two galaxy-halo models using the nearest-neighbor statistic, $d_1$, as a function of $M_{\rm UV,0}$ and $M_{\rm UV,lim}$. Each panel corresponds to a different redshift. The required number of pointings decreases towards higher redshift because the predicted environments of the two models become more distinct, although this requires progressively deeper observations to detect the relevant neighboring galaxies.} 
    \label{fig:ks_all}
\end{figure*}

In the previous sections, we have demonstrated that the environments of UV bright galaxies, specifically the distance to the nearest neighbor, $d_1$, differ between models in which UV luminosity is uniformly boosted at fixed halo mass and those which have high stochasticity in the UV luminosity-halo mass relation.
We now seek to quantify the optimal survey strategies to discriminate between these models.
As described in Section~\ref{sec:obs_choice}, the ability of the nearest-neighbor distance to separate the two models depends on the luminosity of the central bright galaxy, $M_{\rm UV,0}$, the limiting magnitude of neighboring galaxies, $M_{\rm UV,lim}$, and redshift, $z$. 
In this subsection, we determine the combinations of survey depth and bright-galaxy luminosity for which the environmental differences between the models can most easily be used to discriminate between them via the nearest-neighbor distance $d_1$, as a function of redshift. We do this by estimating the number of bright galaxy environments required to significantly distinguish between the two models.

To quantify the observational 
requirements needed to distinguish between the two models, we compare their predicted nearest-neighbor distributions using the Kolmogorov--Smirnov (KS) test. For a fixed combination of $M_{\rm UV,0}$, $M_{\rm UV,lim}$, and $z$, we forward-model the environments of $N$ bright galaxies and compute the resulting distribution of $d_1$ for both models. Following Fig.~\ref{fig:ang_dist}, we set the size of the regions to correspond to a single module of a NIRCam pointing (see Section~\ref{sec:ang}), as the clustering information from the nearest neighbors is concentrated in the nearest $\sim2$\,arcmin. We then determine the minimum number of bright galaxies required to distinguish between the two model distributions at the $1\sigma$ level using the KS test. To reduce Monte Carlo noise, we repeat this procedure $1000$ times and adopt the median value of the number of bright galaxies necessary for the KS test to be passed at $1\sigma$ level\footnote{We also tested a more conservative choice using the $95$th percentile of the distribution, finding the same trends, albeit with a higher normalization.}. We explore $M_{\rm UV,0} \in [-20.5, -22.0]$ \citep[where the upper limit is motivated by possible AGN contamination; e.g.,][]{Kulkarni2019, Finkelstein2022}, $M_{\rm UV,lim} \in [-16.5, -19.5]$  and $z \in \{8, 10.5, 12, 14\}$. The range of $M_{\rm UV,lim}$ spans the UV magnitudes of currently identified photometric neighbors around GN-z11 and GS-z14-0 \citep{Tacchella2023, Robertson2024, Hainline2026}, and extends $\sim1$\,mag fainter than existing photometric limits \citep{Whitler2025, Weibel2026} to explore the benefits of deeper imaging.

Measuring the distribution of nearest neighbor distances, $p(d_1)$, requires that each bright galaxy have at least one photometric neighbor within the survey area.
As our simulations predict that a small fraction of bright galaxies do not have a detectable neighbor, a slightly larger sample is needed to measure $p(d_1)$ than we recover from our KS test.
We account for this by rescaling our resulting required sample size by dividing by the probability of a bright galaxy having at least one detectable neighbor, $p(N(M_{\rm UV,lim}) > 1 \mid M_{\rm UV,0}, z)$.

The absence of a detected neighbor can also be an important constraint. Since the high-stochasticity model associates bright galaxies with lower-mass halos, it predicts a larger fraction of isolated systems compared to the high-luminosity model. For example, at $z=14$, in the high-luminosity model just $<0.8$\% of $M_{\rm UV,0}\approx -21.0$ galaxies are expected not to have a fainter $M_{\rm UV,lim}\lesssim-18$ neighbor, while the high-stochasticity model predicts $\sim5$\% of bright galaxies will have no neighbor in a NIRCam pointing.
Therefore, a NIRCam pointing with no detected photometric candidates at these magnitudes would favor the high-stochasticity model.
Both detections and non-detections of neighboring galaxies thus provide complementary constraints on the galaxy-halo connection \citep[see also][]{Ren2018}. Hereafter, we focus our analysis on the cases \textit{with} detections, as they represent the majority of cases. 

In Fig.~\ref{fig:ks_all} we show the minimum number of bright galaxy environments required to distinguish between our high luminosity and high stochasticity model, at $1\sigma$ significance, as a function of the survey depth ($M_{\rm UV,lim}$) and the luminosity of the bright galaxy ($M_{\rm UV,0}$). We show the results at four different redshifts, $z=8-14$. 
The lightest yellow regions show where the fewest number of bright galaxy environments are required to distinguish the models, and thus the most efficient range to probe observationally. 
In general, we find the models become easier to distinguish as the luminosity of the bright galaxies sampled increases (i.e. for brighter $M_{\rm UV,0}$).
This behavior arises from the tighter halo mass-luminosity relation in the high luminosity model, which produces significantly stronger clustering around UV bright galaxies than in the high stochasticity model. 
As a result, progressively fewer bright galaxy environments are required to distinguish the models as $M_{\rm UV,0}$ becomes brighter.
In the following, we mostly focus on bright galaxies in the range of the most luminous known sources at $z>10$ ($M_{\rm UV,0} \lesssim -21$), and describe our results for the optimal limiting magnitude as a function of redshift.

We begin by describing the optimal parameter range at $z=8$, shown in the upper left panel of Fig.~\ref{fig:ks_all}. We find three regimes that are common to all redshifts, so we describe each for $z=8$ and refer to them at higher redshifts.

At $z=8$, we find the models can be distinguished with a sample of $\sim15-20$ bright $M_{\rm UV,0} \lesssim -21$ galaxies, provided the limiting magnitude of the neighbors is not extremely faint ($M_{\rm UV,lim} \gtrsim-18.7$). This defines three regimes for the limiting magnitude:

\begin{enumerate}
\item \textit{Too-faint} ($M_{\rm UV,lim} \gtrsim -18.7$ at $z=8$): as the limiting magnitude becomes fainter, the number density of neighbors increases, and the nearest neighbor is usually found within the central overdensity in \textit{both} models (i.e., within the central $\sim1$ cMpc; see Fig.~\ref{fig:muvlim_change}, middle panel).
Thus the two models' $d_1$ distributions become increasingly similar, and a larger sample of bright galaxy environments is required to distinguish them.

\item \textit{Optimal} ($M_{\rm UV,lim} \approx -18.7$ to $-19.5$ at $z=8$): for brighter limiting magnitudes, the number density of neighbors decreases and the typical halo mass of the observable neighbors increases.
It becomes less likely that the nearest neighbor lies within $1$ cMpc (the width of the $d_1$ peak in Fig.~\ref{fig:ang_dist}) for the high stochasticity model compared to the high luminosity model, leading to the most distinct $d_1$ distributions between the two models and thus the fewest bright galaxy environments required for discrimination.

\item \textit{Too-bright} ($M_{\rm UV,lim} \lesssim -19.5$ at $z=8$): if the limiting magnitude is too bright, the number density of neighbors becomes so low that neighbors no longer probe the central overdensity hosting the bright source, instead tracing separate density peaks, and the $d_1$ distributions for both models again become similar.
\end{enumerate}

We continue the analysis to higher redshifts in the upper right panel of Fig.~\ref{fig:ks_all}, which shows the results for $z=10.5$. Compared to $z=8$, fewer bright galaxies are required to distinguish between the models, $\sim8$--$15$, provided that the neighboring population is not extremely faint ($M_{\rm UV,lim}\lesssim-18.2$). The reduction in the number of bright galaxies required for model discrimination arises because galaxies of fixed UV luminosity become increasingly rare at higher redshift. In the high luminosity model, the tight connection between UV luminosity and halo mass implies that these rare galaxies reside in correspondingly rare and highly clustered dark matter halos, maintaining small values of $d_1$. 
In contrast, the high stochasticity model allows similarly bright galaxies to arise from lower-mass halos through strong upward fluctuations in UV luminosity. These halos occupy less overdense environments, shifting the $d_1$ distribution toward larger separations. As a result, the difference between the two $d_1$ distributions increases with increasing redshift, reducing the number of bright galaxies needed to distinguish between the models, as we saw in the right panel of Fig.~\ref{fig:muvlim_change}.
As the difference between the distributions becomes larger at fixed survey limit, the optimal regime is thus wider, spanning $M_{\rm UV,lim}\in[-18.0,-19.5]$. 

We show the corresponding results for $z=12$ in the lower left panel of Fig.~\ref{fig:ks_all}. The same overall trend continues, with only $\sim7$--$15$ bright galaxies required to distinguish between the models. 
However, unlike at $z=10.5$, reaching these lower sample sizes requires deeper observations ($-18.5 \lesssim M_{\rm UV,lim}\lesssim -17.5$), as the reduced number densities means that the nearest $M_{\rm UV}\lesssim-18.5$ neighbor is increasingly likely to lie outside the local overdensity in both models, shifting the too-bright regime to lower luminosities.
The reduced number densities also shifts the too-faint regime to lower luminosities, beyond $M_{\rm UV,lim}\gtrsim-16.5$. As the high stochasticity model predicts relatively few close $M_{\rm UV,lim}\lesssim-16.5$ neighbors ($d_1\sim1$ cMpc), while the high luminosity model retains a pronounced peak at small separations, the two models remain well separated over a broad range of survey depths.

At even higher redshift, $z=14$, shown in the lower right panel of Fig.~\ref{fig:ks_all}, we see a similar trend, but the required number of bright galaxies decreases further to $\lesssim5$ (for $M_{\rm UV,0}\lesssim -21.0$), provided observations reach depths of $M_{\rm UV,lim}\gtrsim-18.0$. 
At these redshifts, the halo mass associated with a galaxy of fixed UV luminosity differs substantially between the two models. Because the halo mass function is extremely steep at $z\sim14$, this difference translates into a much larger contrast in the expected environments: in the high luminosity model, bright galaxies occupy exceptionally rare, highly overdense halos, whereas in the high stochasticity model they are typically produced by upscattering in considerably lower-mass halos. Consequently, even a small number of observed environments can place meaningful constraints on the models at $z\sim14$. In particular, the detection of a close photometric neighbor around a bright galaxy (i.e., within $d_1<2$ cMpc) would strongly favor the high luminosity model, while systematically isolated systems would instead support the high stochasticity scenario.

The KS test provides a robust forecast of the sample size required to distinguish the two models. However, for an individual observed system it is often more useful to ask how strongly the observation disfavors one model relative to the other. We therefore also consider a likelihood-based approach, in which we evaluate the probability of data drawn from one model under the probability distribution predicted by the other. Concretely, we evaluate the probability distribution using the kernel density estimate (KDE) of the high luminosity model and draw samples from the high stochasticity model. At $z=10.5$ and $z=12$, we find that even a single galaxy can be sufficient to distinguish the models at $>2\sigma$ over much of the parameter space. This demonstrates that individual systems can already provide meaningful constraints using a likelihood-based approach. However, the likelihood calculation is inherently asymmetric: it assumes one model generates the data and evaluates the probability assigned to those data by the competing model. By contrast, the KS test compares the two distributions directly without treating either as the reference model. For this reason, we use the KS test as our primary forecasting tool, while the likelihood calculation provides intuition for the constraining power of individual systems. 

To summarize, distinguishing between the models at $z=8$ requires observations of a relatively large number of bright galaxies ($M_{\rm UV,0}\lesssim -21.0$, $N\sim18$), although only moderate depth is needed to establish their neighbor population ($M_{\rm UV,lim}\approx-19$). At $z=10.5$, fewer bright systems are required ($N\sim10$), with limiting magnitudes comparable to current JWST surveys, $M_{\rm UV,lim}\sim-18.5$ \citep{Tacchella2023, Whitler2025}. Similar sample sizes of UV bright galaxies are sufficient at $z\sim12$, though deeper observations of their neighbors are needed ($M_{\rm UV,lim}\approx-18$). At the highest redshifts, $z\sim14$, we find only a small number ($\lesssim5$) of bright galaxies is required, provided observations reach depths of $M_{\rm UV,lim}\approx-18$. Overall, the optimal observational strategy changes with the targeted cosmic epoch, with deeper observations needed for increasing redshift.
In Section~\ref{sec:future} we discuss the required survey areas to build these samples of bright $z>8$ galaxies.

\section{Comparison to observations} 
\label{sec:obs}
The nearest-neighbor statistic introduced in Section~\ref{sec:methodology} is particularly well suited for application to the brightest spectroscopically confirmed galaxies at $z>10$. In this section, we apply the framework to the environments of GN-z11 (Section~\ref{sec:gnz11}) and GS-z14-0 (Section~\ref{sec:gsz14}) to assess whether their observed neighboring populations favor the high luminosity or high stochasticity scenario.
In the following, we compare these observations to the nearest-neighbor distance statistic introduced in Section~\ref{sec:ang}.

\subsection{The environment of GN-z11}
\label{sec:gnz11}

\begin{figure*}
    \centering
    \includegraphics[width=1.0\linewidth]{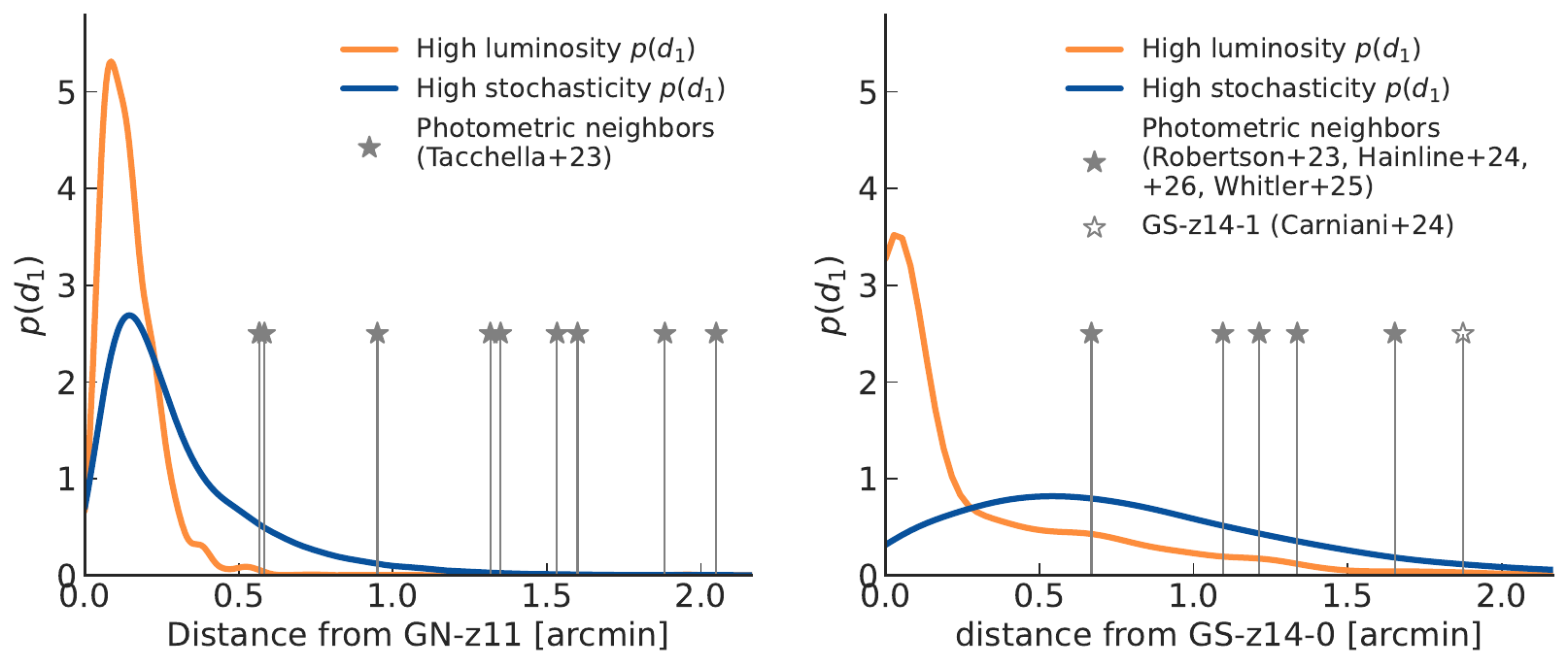}
    \caption{Probability distribution function of distances to the closest neighbor of the brightest galaxies at $z>10$ for the high luminosity (orange) and high stochasticity (blue) models. \textit{Left panel:} Predictions of the two models for the environment of GN-z11 at $z=10.6$ (using our fiducial line-of-sight window of $\Delta z=1$; Section~\ref{sec:gnz11}). All neighbors of GN-z11 from \citealt{Tacchella2023} are shown. The closest neighbor lies at a larger distance than predicted by the high luminosity model, implying a preference for the high stochasticity scenario. \textit{Right panel:} Predictions of the two models for the environment of GS-z14-0 at $z=14.1$ (using $\Delta z=1$; Section~\ref{sec:gsz14}). The photometric candidates within $2$ arcmin from \citet{Robertson2023}, \citet{Hainline2024}, \citet{Hainline2026} and \citet{Whitler2025} are shown in gray. All of the photometric candidates are found at a larger distance than $0.5$ arcmin, implying a preference for the high stochasticity scenario.}
    \label{fig:gnz11_gsz140}
\end{figure*}

GN-z11 is the brightest known galaxy at $z > 10$, making it a prime target for analysis of its environment. \citet{Tacchella2023} identified nine photometric candidate galaxies within 10 cMpc (corresponding to $3.5$
\,arcmin). The candidates span a UV-magnitude range of $-19.9 < M_{\rm UV} < -18.2$, enabling a direct probe of GN-z11's local environment. GN-z11 has also been suggested to host an AGN \citep{Bunker2023, Maiolino2024, Scholtz2024, Chen2026b}; throughout this section we treat it as a star-forming galaxy, using its $M_{\rm UV}$--$M_{\rm h}$ relation as in Section~\ref{sec:analytical_framework}, and return to the implications of a possible AGN contribution at the end of this section.

In the left panel of Fig.~\ref{fig:gnz11_gsz140}, we show the angular separation of the photometric neighbors from GN-z11 as gray stars, along with the predicted distribution of the nearest-neighbor separation for our two models from Section~\ref{sec:methodology}. 
The model predictions (shown with colored curves) are matched to the area and depth of the JADES observations described by \citet{Tacchella2023}, including the size of the field around GN-z11 ($L =10$ cMpc, or $3.5$\,arcmin), redshift ($z=10.6$, and a fiducial line-of-sight window of $\Delta z\approx1$,
UV luminosity of the bright galaxy ($M_{\rm UV,0} = -21.0$) and the limiting magnitude of the neighbors ($M_{\rm UV,lim} = -18.2$) (see also Sec.~\ref{sec:methodology}). 
We note the candidates' best-fit photometric redshifts reported by \citet{Tacchella2023} span slightly wider than $\Delta z\approx1$. 
Our narrower window reflects the assumption that photometric redshift uncertainties place the true redshifts of these candidates closer to GN-z11 than their best-fit photometric redshift solutions suggest. 
However, as we discuss below, using a wider redshift window does not impact our conclusion.

We now compare the observed separations to the model predictions.
In the high luminosity model, there is a 98\% probability that the nearest neighbor should be closer than 0.5\,arcmin.
Strikingly, none of the observed candidates from \citet{Tacchella2023} are within that angular separation. The galaxy closest to GN-z11 is 0.56\,arcmin from GN-z11.
In contrast, the high stochasticity model predicts a broader distribution of separations with an extended tail: there is a 20\% probability of finding neighbor $>0.5$\,arcmin from a galaxy as bright as GN-z11. Thus, the distance to GN-z11's nearest neighbor appears more consistent with the high stochasticity model.

To quantify the preference for the high stochasticity model, we evaluate the likelihood of finding the nearest neighbor at $0.56$\,arcmin for each model by computing the probability distribution at that location. We find that the high stochasticity model has a likelihood $7\times$ higher than the high luminosity model: corresponding to $\Delta \ln \mathcal{L} \approx 1.97$, or roughly a $\sim 2.0 \sigma$ preference (assuming that the likelihood-ratio statistic follows a $\chi^2$ distribution with one degree of freedom, \citealt{Wilks1938}). The likelihood ratio thus suggests GN-z11 may reside in a relatively low mass halo, with its luminosity boosted by a recent burst of star formation. A strong recent burst is consistent with the star formation history inferred from the NIRCam SED \citep{Tacchella2023}, and the detection of massive star stellar wind features in the ultra-deep SPURS spectrum \citep{Chen2026b}. 
Repeating the analysis with a wider $\Delta z=2$ window, the high stochasticity model remains preferred, at the $\Delta\ln\mathcal{L}\approx1.67$ level ($\sim1.8\sigma$).

We can obtain a rough estimate of the host halo mass for GN-z11 implied by our high stochasticity model, though a robust mapping to halo mass will ultimately require marginalizing over the UV magnitude–halo mass relation with a wider range of models than the two considered here.
We select the halos hosting $M_{\rm UV,0}<-21.5$ galaxies in the high stochasticity catalog which have their nearest neighbor at $d_1>0.5$\,arcmin. We find these halos have masses $\log (M_{\rm h}/M_{\odot})\in[10.1,10.6]$ ($68\%$ C.I.). 
By contrast, in the high luminosity model, galaxies with the same $M_{\rm UV,0}$ are hosted in much more massive halos: $\log (M_{\rm h}/M_{\odot})\in[11.0, 11.3]$.

The nine galaxies identified around GN-z11 are currently photometric candidates, and spectroscopic confirmation will be required to establish their redshifts and true three-dimensional separations from GN-z11, as well as to rule out possible low-redshift interlopers. 
However, these neighbors have photometric redshift solutions that are highly consistent with that of GN-z11, making them likely physically associated systems. Importantly, we note that if the closest galaxy is found to be at a different redshift (corresponding to more than $\approx10$\,cMpc in the line-of-sight direction) from GN-z11 this would shift the observed nearest-neighbor distance to larger values, increasing the preference for the high stochasticity model, strengthening our conclusion.

We note that JWST/NIRSpec spectroscopy of GN-z11 has indicated possible signatures of AGN activity \citep{Bunker2023, Maiolino2024, Scholtz2024,Chen2026b}.
If part of the observed UV luminosity of GN-z11 originates from an AGN, this can provide an additional source of scatter in the relation between UV luminosity and halo mass, effectively increasing the stochasticity already captured in the high-stochasticity model. Therefore, the combination of GN-z11's extreme UV luminosity and relatively isolated environment remains consistent with the high stochasticity model. 
While AGN feedback may suppress star formation in close neighbors, potentially reducing the number of nearby companions \citep{Morganti2017}, given the likely low luminosity of an AGN in GN-z11, this effect is expected to be confined to scales of a few kpc \citep{Maiolino2026} and is thus unlikely to significantly impact the larger-scale environment probed here.

\subsection{The environment of JADES-GS-z14-0}
\label{sec:gsz14}

We now seek to apply our framework to
JADES-GS-z14-0 (or GS-z14-0), the most luminous galaxy ($M_{\rm UV} \approx -20.8$) spectroscopically confirmed at $z>14$ \citep[$z_{\rm spec}=14.18$,][]{Carniani2025, Carniani2026}. 
The number density of $z\sim14$ galaxies, including the most luminous confirmed sources GS-z14-0 and MoM-z14 \citep[$z_{\rm spec}=14.44$,][]{Naidu2025}, poses the biggest challenge to galaxy formation models, making clustering estimates at $z\sim14$ critical.

Here, we consider candidate photometric neighbors of GS-z14-0 in the GOODS-S field, selected from the JADES survey, previously reported by \citep{Robertson2024, Whitler2025, Hainline2024, Hainline2026}, and compare their angular separations from GS-z14-0 to our model predictions.
We select galaxies located within $2$ arcmin of GS-z14-0 whose $90\%$ photometric-redshift confidence intervals overlap with the spectroscopic redshift of GS-z14-0 ($z_{\rm spec} = 14.32$), starting from the parent photometric catalogs of \citet{Robertson2024, Whitler2025, Hainline2024, Hainline2026}. This mimics the selection of neighbor candidates of GN-z11 from \citet{Tacchella2023}. This yields a sample of six photometric candidates. Spectroscopic follow-up of two of these sources was presented by \citet{Carniani2024}. One galaxy, GS-z14-1, was confirmed to lie at a similar redshift ($z_{\rm spec}\approx13.9$, shown with a white-filled star in the right panel of Fig.~\ref{fig:gnz11_gsz140}), while the other source, JADES-GS+53.10763-27.86014, was not yet robustly detected spectroscopically. The reported neighboring candidates span a UV-magnitude range of $-18.6 < M_{\rm UV} < -18.0$. The observations are likely complete for these UV magnitudes at $z\sim 14$ \citep{Whitler2025}.

To compare our framework to these observations, we construct mock realizations from our $z=14$ catalogs, as detailed in Section~\ref{sec:red_dep}. We extract nearest-neighbor separation distributions, $p(d_1)$, from our catalogs by selecting $L = 2\,\mathrm{arcmin}$-sized regions (corresponding to a single module of a NIRCam pointing, as motivated in Section~\ref{sec:ang}) around bright galaxies ($M_{\rm UV,0}=-20.8$), applying a limiting magnitude for neighbors of $M_{\rm UV,lim}<-18.0$. As for GN-z11 (Section~\ref{sec:gnz11}), we adopt $\Delta z=1.0$ as a fiducial line-of-sight window\footnote{As at GN-z11, GS-z14-0's candidate neighbors have best-fit photometric redshift solutions spanning closer to $\Delta z\approx2$; however, our conclusions are unchanged adopting the wider range.}. Our simulations provide a sample of $\sim70,000$ bright galaxy environments for this analysis.

In the right panel of Fig.~\ref{fig:gnz11_gsz140} we show the angular separation of the candidate neighbors from GS-z14-0, along with the predicted distribution of separations to the closest neighbor from our two models.
The closest candidate neighbors reported in the three catalogs are separated by $0.58-1.3$\,arcmin from GS-z14-0; the exact value depends on which catalog is used, as they differ in which sources are counted as candidates and in their estimated redshifts, so the identity of the true nearest neighbor is ambiguous.
Consistent with expectations from lower redshifts, the high luminosity model predicts a much higher probability of the nearest neighbor being within 0.5\,arcmin of GS-z14-0 ($80\%$ of samples) than the high stochasticity model ($34\%$ of samples).
While the tail to larger separations increases in the high luminosity model at $z\sim14$ relative to $z\sim10$, due to decreasing number densities (see Section~\ref{sec:red_dep}), we find that the high stochasticity model is still preferred at the $\Delta \ln \mathcal{L}\approx 0.83-0.94$ level (or roughly a $\sim 1.1 \sigma$ preference). 
As above, this range reflects the same catalog-dependent choice of nearest neighbor. Again, we note that if spectroscopic follow-up excludes the closest candidates as $z\sim14$ sources, this will strengthen our conclusion.

As for GN-z11, we can estimate the halo mass of GS-z14-0 implied by each model. We find that the high stochasticity model predicts $M_{\rm h}\in[8.8, 10.1]$ (68\% C.I.), while the high luminosity model predicts $M_{\rm h}\in[10.4, 10.7]$. We again select bright galaxies ($M_{\rm UV,0}<-20.6$) with nearest neighbors at $d_1>0.5$,arcmin. We find that the resulting halo-mass distributions do not shift significantly after this additional conditioning. Although the large nearest-neighbor separation provides some discrimination between the two models, it does not substantially constrain the halo mass of GS-z14-0. Deeper observations to detect fainter neighbors will therefore be required to place stronger constraints on the halo mass of GS-z14-0 (see Section~\ref{sec:future}).

The results in this section are based solely on the nearest-neighbor statistic, $d_1$. Incorporating the spatial distribution of all of the neighboring galaxies would likely provide stronger discriminatory power between the models. In addition, applying the framework to larger samples of bright galaxies across multiple redshifts would substantially improve the statistical constraining power of the method (see Section~\ref{sec:ks_test}). We aim to explore both avenues in future work as we discuss in Section~\ref{sec:future}.

\section{Discussion}
\label{sec:discussion}

The unprecedented population of $z>10$ galaxies discovered by JWST has motivated significant theoretical efforts to understand the primary mechanism for their abundance, though distinguishing between proposed models requires information beyond current UVLFs. In this paper, we have investigated clustering statistics that can help distinguish between different classes of these models.
We have shown the nearest-neighbor clustering statistic can be a powerful way to discriminate between galaxy formation models at $z>10$. In the previous section we demonstrated current observations of the brightest galaxies at $z>10$ favor a scenario in which a high amplitude of stochasticity plays an important role in shaping the UV-bright galaxy population at $z>10$. 
In this section, we place this result in the context of the on-going debate about the origin of the $z>10$ UVLF excess (Section~\ref{sec:disc_consensus}).
We also discuss future observational prospects for using the framework introduced here, and possible extensions to the method, to further constrain the galaxy-halo connection at cosmic dawn (Section~\ref{sec:future}).

\subsection{The importance of stochasticity at $z>10$} \label{sec:disc_consensus}

Our analysis of the environments of GN-z11 and GS-z14-0 (Section~\ref{sec:obs}) suggests that stochasticity is important for explaining their visibility.
If the $z>10$ UVLF excess was primarily driven by a uniform boost in UV luminosity at fixed halo mass, we expect these UV-bright sources to trace the most massive halos, with neighbors within $<0.5$\,arcmin. Instead, our results indicate that, at these epochs, UV-bright galaxies may not be exclusively hosted in the most massive dark-matter halos.

From a theoretical perspective, stochasticity is generally expected to become more important at higher redshifts, as star formation shifts to progressively lower mass halos. Low mass halos have shallower potential wells and fewer star-forming regions that provide stable star formation rates in more massive halos \citep{Sparre2017, Ma2018, Gelli2024}.
At fixed halo mass, star formation may even become burstier towards higher redshifts: increasing gas density implies the timescales of the dominant feedback mechanisms, such as supernova energy injection, may be comparable to the dynamical times of the system, which prevents self-regulation and renders star formation stochastic \citep{Faucher-Giguere2018, Pallottini2024, Sun2025}. 
However, the precise source and strength of feedback in high-redshift galaxies remain debated, and simulations do not converge: subgrid prescriptions tuned to a given observable (e.g.\ the UVLF) can yield markedly different predictions for individual galaxy properties \citep{Ma2018, Sun2023, Pallottini2024, Jeong2025, Li2024, Feldmann2025}, making observations beyond the UVLF particularly important.

JWST's ability to constrain the rest-UV to optical SEDs and spectra for $z\gtrsim6$ galaxies provided our first hints of large stochasticity in the galaxy-halo connection at high redshift.
At fixed UV magnitude, a large variance has been reported in nebular emission line strengths \citep{PrietoLyon2023,Endsley2023,Endsley2024} and SED shapes \citep{Ciesla2024, Carvajahal-Bohorquez2025, Cole2025} at $z\sim4-9$, indicating we are capturing galaxies at different phases of their star formation histories.
Notably, \citet{Endsley2024} demonstrated that the $H\alpha$-to-UV flux ratio, tracing the ratio of recent ($\sim10$\,Myr) to longer-timescale ($\sim100$\,Myr) star formation, broadens at fainter UV magnitudes at $z\sim4-6$. \citet{Munoz2026} fit these observed $H\alpha$-to-UV distributions, inferring that star formation stochasticity increases toward lower halo mass. Their inferred $\sigma_{\rm UV}\sim0.75$--$1.5$\,mag at $\log M_{\rm h}/M_\odot\sim11$--$9$ is consistent with the $\sigma_{\rm UV}\sim0.9$--$1.6$\,mag predicted by our high stochasticity model over the same halo mass range.
If stochasticity increases towards higher redshifts, as expected due to halo mass-dependent stochasticity, we expect both a larger number of galaxies caught in SFR upturns -- visible as strong emission-line sources -- and a corresponding population of weak emission-line sources caught in recent downturns with increasing redshift (Gelli et al. in prep).
The first statistical samples of $z>9$ galaxies' spectra have demonstrated that sources with strong emission lines are more common compared to $z\sim6$, indicating a growing population of galaxies undergoing significant SFR upturns \citep{Roberts-Borsani2025, Tang2026}, and with deep UV spectroscopy revealing star formation in extreme high density, low metallicity conditions \citep{Chen2026b}. 
Strikingly, detailed studies of individual $z>9$ galaxies with weak emission lines demonstrate these sources are consistent with being observed in a downturn following a recent burst of star formation \citep{Chen2026, Harikane2026}.
Together, the spectral diversity of $z>6$ galaxies indicates significant stochasticity in star formation histories. 

JWST has also enabled the first estimates of galaxy clustering at $z\sim6-10$ via angular correlation functions \citep{Paquereau2025, Shuntov2025,Dalmasso2024,Dalmasso2026} and pure-parallel cosmic variance statistics \citep{Weibel2025,Kreilgaard2026}. 
These studies broadly agree that UV-bright galaxies are more strongly clustered than UV-faint galaxies at these redshifts, implying a correlation between UV luminosity and halo mass, but the role of stochasticity remains debated.
Indeed, both \citet{Weibel2025} and \citet{Kreilgaard2026} find that the number densities of bright $z\sim7-10$ galaxies are difficult to reconcile with the clustering inferred from their cosmic variance, considering models both with and without stochasticity, potentially pointing to missing ingredients in the interpretation of cosmic variance clustering estimates, for example non-linear effects or cosmological assumptions.
At the same time, angular correlation function studies have also provided evidence that stochasticity is important for explaining galaxy clustering.
In particular, \citet{Shuntov2025} measured the angular correlation function at $z=3.8$--$9.0$ for a H$\alpha$- and [OIII]-selected sample, complete down to $M_{\rm UV}\lesssim-19.1$. 
Using HOD modeling they inferred a scatter in the $M_{\rm UV}$--$M_{\rm h}$ relation of $\sigma_{\rm UV}\approx0.7\pm0.3\,$mag (assuming no mass dependence), and found no significant evolution over this redshift range. This is significantly higher than the $\sigma_{\rm UV}\approx0.3\,$mag predicted from stochasticity in halo accretion rates alone \citep{Ren2018}, indicating additional stochasticity in star formation. 
This result is consistent with our high-stochasticity model if we consider the same redshift and $M_{\rm UV}$ range: we predict $\sigma_{\rm UV}(M_{\rm UV}=-19.1, z=8)\approx0.87\,$mag (see Sec.~\ref{sec:red_dep}).
Our analysis provides the first constraints on clustering at $z>10$, suggesting that stochasticity may play an increasing role at higher redshifts.

Determining the evolution of the galaxy-halo connection at $z>10$ will require progress on both observational and theoretical fronts. 
Most importantly, this motivates self-consistent clustering estimates across a broader redshift range, including $z>10$, so that redshift evolution can be properly characterized.
The nearest-neighbor statistics developed here offer one route to this, requiring fewer sources than angular correlation functions to estimate clustering.
Spectroscopic confirmation of photometric candidates will also be important: while our nearest-neighbor clustering statistic is robust to interlopers, even a single interloper can significantly inflate cosmic variance clustering measurements. 
On the theoretical side, improved modeling of non-linear and small-scale bias \citep{Wang2026} and further exploration of the sensitivity of these measurements to cosmological parameter assumptions will be important for understanding the theoretical systematics.
Ultimately, deep imaging of more bright $z>10$ galaxies will be essential for realizing the potential of JWST to constrain the galaxy-halo connection at our redshift frontier, which we discuss below.

\subsection{Future prospects}
\label{sec:future}

Corroborating the high stochasticity implied by our analysis of GN-z11 and GS-z14-0 in Section~\ref{sec:obs} will require building a larger sample of bright galaxies' environments (see Section~\ref{sec:ks_test}).
Achieving this requires both deeper imaging around known bright sources, to detect their neighbors, and wider area imaging to build a larger sample of bright $z>10$ galaxies.
In this section, we discuss both avenues, as well as extensions of our framework which should improve the constraining power of individual pointings.

We demonstrated in Section~\ref{sec:obs} that even observations of the environments of single sources can place constraints on galaxy formation models. As we discussed in Section~\ref{sec:obs_choice}, the tightest constraints will come from observations which sample the optimal magnitude range to distinguish between the models, as well as spectroscopic confirmation of the faint neighbors to establish their physical association (or not) to the bright galaxy.
For the two sources we discussed in this work, additional observations would thus improve the constraints we presented.
For GN-z11, spectroscopic confirmation of its faint candidate neighbors would establish their physical association with GN-z11 -- ruling out the nearest photometric candidates would further favor the high stochasticity model.
For GS-z14-0, deeper imaging reaching $M_{\rm UV,lim}\sim[-17,-18]$ would probe the fainter neighbors that maximize the difference between the high-luminosity and high-stochasticity models (Section~\ref{sec:ks_test}).

The same depth requirement applies to other bright $z>10$ galaxies confirmed by JWST, which still lack deep imaging of their surroundings.
For example, deep NIRCam imaging around the bright $z\sim12$ source GHZ2/GLASS-z12 \citep[$M_{\rm UV}=-20.5, z=12.34$,][]{Castellano2024} is still uneven as it lies at the edge of the Abell 2744 field \citep{Atek2023, Castellano2023}, and no candidate neighbors have yet been reported within $2$ arcmin.
Similarly, at $z\sim13-14$, PAN-z14-1 \citep[$M_{\rm UV}=-20.6, z=13.53$,][]{Donnan2026} and MoM-z14 \citep[$M_{\rm UV}=-20.2, z=14.44$,][]{Naidu2025} both lack imaging deep enough to detect their faint neighbors \citep[existing limiting magnitudes are $M_{\rm UV}\lesssim-19, -20$ in these fields respectively,][]{Donnan2024,Williams2025}.
Our simulations in Section~\ref{sec:obs_choice} demonstrate that imaging reaching $M_{\rm UV}\sim -18$ mag covering $R < 2$\,arcmin around these sources will provide good discriminatory power for our models.
This would be possible in just a single NIRCam pointing in $\approx11$--$13$ hours per broad-band filter (for $5\sigma$ detections).

Establishing larger samples of bright $z>10$ galaxies, requiring wide area surveys, will also improve the clustering constraints explored in this work, and better enable measurements of redshift evolution. 
In Section~\ref{sec:ks_test} we showed that at $z=10-12$, $\sim8-15$ bright galaxies ($M_{\rm UV,0}\lesssim-20.5$) are required to distinguish between the high luminosity and high stochasticity models at high significance. Using our model UV luminosity functions and accounting for Poisson noise (requiring a conservative $>95\%$ probability of detecting the required number of galaxies), we find that building this sample would require large survey areas of $2.2$ ($3.5$) deg$^2$ at $z=10.5$ ($z=12$).
At $z=14$, we found fewer bright galaxies are needed, $\approx5$, requiring a smaller area of $\approx1.5$ deg$^2$.
The area of current JWST imaging suitable for $z>10$ galaxy selection is still below these requirements \citep[$\sim 0.6$ deg$^2$,][]{McLeod2026}; however, the prospects for building larger samples are promising.
At $z\sim10-12$, detecting sufficient numbers of bright galaxy candidates ($M_{\rm UV,0}\approx-20.5$) should be feasible with wide-area NIR deep imaging surveys with Roman \citep[][]{RomanWFI2025} and Euclid \citep{EUCLID2025}. 
For example, the deep tier of the Roman High Latitude Wide Area Survey will observe $\approx20$\,deg$^2$, reaching a limiting magnitude of $M_{\rm UV}\approx-20.0$ ($m_{AB}\approx27.5$) at $z\approx10$ (F158, $5\sigma$) \citep{Roman2025}.
Furthermore, Euclid proposed deep surveys will cover a much wider area ($\approx 50$\, deg$^2$), at a depth corresponding to $M_{\rm UV}\approx -21.0$ ($m_{AB}\approx26.5$) at $z\approx 10$ \citep[][]{McPartland2025, EUCLID2025}.
The synergy of bright galaxy selection using Roman and Euclid with deep imaging with NIRCam to investigate their environments (and spectroscopic confirmation with NIRSpec), promise to rapidly improve our understanding of luminous $z>10$ galaxies.
At higher redshifts, continued efforts to build wider areas with deep $>2$\,micron imaging will be essential to identify robust bright $z\gtrsim12$ candidates.

While in this work we have focused on the position of only the closest neighbor, future work to include more information about the bright galaxies' neighbors should further improve our ability to distinguish galaxy formation models.
In particular, jointly modeling the distribution of multiple neighboring galaxies' separations would improve constraints by including additional information about the 3D density distribution.
While there are non-trivial covariances between the positions of individual galaxies, making an analytic joint likelihood approach challenging, 
this could be done using simulation-based inference (SBI), where full cutouts can be forward-modeled, allowing us to efficiently use simulations and data to generate non-Gaussian likelihoods \citep{Ho2024, deSanti2025, Fischbacher2025}. 
Furthermore, including other galaxy properties, 
such as the distributions of nebular lines equivalent widths, that independently constrain the stochasticity of star formation histories \citep[e.g.,][Gelli et al., in prep]{Endsley2023, Endsley2024, Munoz2026}, could enable us to capture more complex galaxy formation models and further constrain star-formation processes at $z>10$. We aim to explore both of these avenues in future work.

\section{Conclusions}
\label{sec:conclusions}

JWST's extraordinary capability to detect galaxies deep into the Cosmic Dawn has revealed a surprising number of bright $z>10$ galaxies; however, the physical cause of their high abundances cannot be determined from the UVLF alone.
Galaxy clustering provides a powerful tool to break degeneracies between galaxy formation models by probing the UV luminosity-halo mass relation, but traditional clustering measurements become very uncertain at $z>10$ due to low number densities.
In this work, we use a forward-modeling framework for the 3D distribution of neighbors around bright $z>10$ galaxies to develop new clustering statistics that can be used to discriminate between theoretical models at the highest redshifts.
Our main findings are as follows:

\begin{enumerate}
    \item We develop an analytical formalism based on conditional mass functions that predicts the number of faint neighbors around a bright galaxy, and their radial separation. This framework provides physical intuition for how clustering signatures at $z > 10$ depend on the galaxy-halo connection. Models which have high stochasticity in UV luminosity at a given halo mass predict bright ($M_{\rm UV}\sim -21$) galaxies located in a wide range of halo masses and overdensities, with neighbors located at larger separations compared to models where the UV luminosity-halo mass connection is tighter.

    \item We build forward models of 3D galaxy distributions, designed to reproduce JWST observations using semi-numerical 21cmFAST simulations, assuming two representative models constructed to match the UVLFs at $z>10$. The ``high luminosity'' model assumes a tight luminosity-halo mass relation, with an increased amplitude at higher redshifts relative to pre-JWST predictions, characteristic of models invoking e.g., a higher star formation efficiency, top-heavy IMF or reduced dust attenuation at $z>10$. The ``high stochasticity'' model invokes a high amplitude of mass-dependent scatter between UV luminosity and halo mass, reflecting predictions for very bursty star formation in low mass halos.
    We use these simulations to characterize the environments of bright galaxies at $z>10$.

    \item We find that number counts of photometric neighbors around bright galaxies are not sufficient to distinguish between the two extreme models at $z>10$, due to large uncertainties from Poisson and cosmic variance. As an example, we find the number of photometric neighbors around GN-z11 is equally well explained by both of our models, motivating additional statistics.

    \item We identify the separation to the closest neighbor of a bright source as a key statistic to discriminate between the two models, as it probes the local clustering.
    We demonstrate both 3D and angular separations provide discriminatory power, meaning our framework can be applied to photometric data. We identify that single NIRCam pointings (corresponding to $2$\,arcmin per side) suffice to probe the clustering around bright galaxies.

    \item We quantify the required number of bright galaxies that need to be targeted and survey depth to optimally distinguish between the two models.    
    The nearest-neighbor separation depends on the UV magnitudes of both the bright galaxy and its neighbors, as well as redshift. 
    At $10 \lesssim z < 12$, we find that targeting $\sim10$ $M_{\rm UV} < -20.5$ galaxies and reaching limiting magnitudes of their neighbors down to $M_{\rm UV,lim}\approx-18.0$ will suffice to distinguish the two models at $68\%$ confidence.
    At $z\sim14$, deeper imaging down to $M_{\rm UV,lim}\sim-17.5$ is needed, though only for $3-5$ bright galaxies.

    \item We apply our framework to two of the brightest spectroscopically confirmed galaxies at $z>10$: GN-z11 \citep{Oesch2016,Bunker2023} and JADES-GS-z14-0 \citep{Carniani2024}, which both have many photometric candidates nearby -- 9 within $2$\,arcmin of GN-z11 \citep{Tacchella2023}, and 6 within $2$\,arcmin of JADES-GS-z14-0 \citep{Robertson2023, Hainline2024, Hainline2026, Whitler2025}.
    The high luminosity model predicts that the nearest neighbor should be within $<0.5$\,arcmin; however, no photometric neighbors are found that close to either source.
    We find the separations to the nearest photometric neighbors are more likely to be consistent with the high stochasticity model, where the bright galaxies are more likely to reside in less massive halos that experience a temporary boost in UV luminosity during, and shortly after, a star formation burst. This suggests that stochasticity, rather than enhanced UV luminosity alone, is important for explaining the luminosity of GN-z11 and JADES-GS-z14-0.

\end{enumerate}

These results provide independent evidence for the importance of star formation stochasticity at $z>10$, based on environmental constraints, complementing UVLF and spectroscopy-based studies. 
Future work to extend this analysis to larger samples, and to improve the discriminatory power between models by incorporating additional information, promise to deepen our understanding of star formation at the earliest epochs.

\begin{acknowledgements}
We thank Lily Whitler, Jordan Mirocha, Michele Trenti, Julian B. Muñoz, Andrei Mesinger and Dan Stark for helpful discussion in preparation of the manuscript. IN acknowledges support by European Union ERC grant RISES (101163035). CAM acknowledges support by the European Union ERC grant RISES (101163035), Carlsberg Foundation (CF22-1322), and VILLUM FONDEN (37459). The Cosmic Dawn Center (DAWN) is funded by the Danish National Research Foundation under grant DNRF140. Views and opinions expressed are those of the author(s) only and do not necessarily reflect those of the European Union or the European Research Council. Neither the European Union nor the granting authority can be held responsible for them. The Tycho supercomputer hosted at the SCIENCE HPC center at the University of Copenhagen was used for supporting this work.
\end{acknowledgements}

\bibliographystyle{aa}
\bibliography{neighbors}

\begin{appendix}

\section{Analytic formalism}
\label{sec:appB}

In this appendix, we derive the analytic framework used to connect the environments of bright galaxies to their UV luminosities. In particular, we develop expressions for the expected number density and spatial distribution of neighboring galaxies conditioned on the presence of a bright source. The derivations presented here motivate the expressions introduced in Sec.~\ref{sec:analytical_framework}.

\subsection{Excursion set basics}

The abundance of dark-matter halos, that serve as birthplaces of galaxies, can be derived using the theory of structure formation \citep[e.g.,][]{Padmanabhan1993}. In particular, the excursion-set formalism developed in \citet{Press1974} provides a starting point for discussing galaxy clustering. A key quantity in this framework is the collapsed fraction of dark matter in halos ($f_{\rm coll}$). Its derivative with respect to mass can be written in terms of the variance of the matter density field ($\sigma_M$), and the critical collapse threshold ($\delta_c$). Namely, within the Press-Schechter formalism we have:

\begin{equation}
    \frac{\textrm{d}f_{\rm coll}(>M,z)}{\textrm{d}M} = \sqrt{\frac{2}{\pi}} \frac{\delta_c(z)}{\sigma_M^2(M)} \left| \frac{\textrm{d}\sigma_M(M)}{\textrm{d}M}\right| \exp\left[ -\frac{\delta_c^2(z)}{2\sigma_M^2(M)}\right]
    \label{eq:def_hmf}
\end{equation}

The collapsed fraction can be used to calculate a quantity more commonly used, the halo mass function:

\begin{equation}
    \frac{{\rm d}n}{\textrm{d}M } \equiv \frac{\overline{\rho}_0(z)}{M} \frac{\textrm{d}f_{\rm coll}(>M, z)}{\textrm{d}M},
\end{equation}

where $\overline{\rho}(z)$ is the mean matter density of the Universe at some redshift $z$. Within this formalism, we write the average number of halos above some mass $M_{\rm min}$ in some volume $V$ as:

\begin{equation}
    N_{\rm halo} (z, V) = V \cdot \int_{M_{\rm min}}^{\infty} \frac{{\rm d}n}{\textrm{d}M } \textrm{d}M.
    \label{eq:standard_number}
\end{equation}

To later connect halo abundances to observable galaxy populations, it is useful to rewrite Eq.~\ref{eq:standard_number} in terms of a general observability function, which defines the fraction of halos with mass $M$ that are selected:

\begin{equation}
    N_{\rm halo} = V \cdot \int dM \frac{dn}{dM} \cdot p_M(M),
    \label{eq:n_gal_norm}
\end{equation}

where $p_M(M)$ is the observability of halos with halo mass $M$. In the simplest case in which we select all of the halos above a given mass,  we recover Eq.~\ref{eq:standard_number} with the Heaviside theta function $p_M(M) = \Theta(M-M_{\rm min})$.  Crucially, we can use Eq.~\ref{eq:n_gal_norm} to extend this formalism to count the galaxies instead of halos. Since galaxies trace the halo abundance, the only difference will be in the form and interpretation of the function $p_M(M)$. For example, if we are quantifying galaxies based on their UV magnitudes, we can write the probability $p_M(M)$ as:
 
\begin{equation}
    p_M(M) = \int \textrm{d}M_{\rm UV} \cdot p(M_{\rm UV}|M) p_{M_{\rm UV}}(M_{\rm UV})
    \label{eq:pM_first}
\end{equation}

This equation tells us that the probability of observing a galaxy hosted by a halo of mass $M$ can be marginalized using a probabilistic connection of the UV magnitude of a galaxy and its halo mass. In the previous equation, $p(M_{\rm UV})$ is the observability of a galaxy with given $M_{\rm UV}$, which is in general a complex function that depends on the galaxy survey and redshift. In this work, we will assume it to be a Heaviside function, and comment on potential caveats from it. Using this, we can differentiate with respect to $M_{\rm UV}$ and recover the UV luminosity function, i.e., Eq.~\ref{eq:standard_uvlf}. 

\subsection{Conditional mass function}
\label{sec:cmf}
The previous subsection described the average abundance of halos in the Universe. However, galaxies form in environments with different large-scale overdensities, implying that the local halo abundance varies spatially. To describe galaxy environments around bright sources, we therefore require the conditional mass function, that describes spatial modulation of the halo mass function within a spherical region of radius $R$ and overdensity $\delta$.
The conditional mass function can be constructed by replacing the condition for collapse in Eq.~\ref{eq:def_hmf}: $\delta_c$ to $\delta_c - \delta$ and mass variance from $\sigma^2$ to $\sigma_M^2 - \sigma_M^{2}(R, z, \delta)$ \citep[see][]{Bond1991}.

We can again write the number of halos in a spherical region of volume $V$ as :
\begin{equation}
    N_{\rm halo}(V, z) = V\cdot \int dM \int\textrm{d}\delta\cdot  \frac{\textrm{d}n(>M, z|\delta, M_{\rm vol})}{\textrm{d}M} \cdot p(\delta),
    \label{eq:ngal_def}
\end{equation}

where $\delta$ represents the linear overdensity over a region of volume $V$ at redshift $z$. $\frac{\textrm{d}n(>M, z|\delta, V)}{\textrm{d}M}$ is the conditional mass function that quantifies the number of halos in a region with overdensity $\delta$. Finally, $M_{\rm vol}$ is the total mass enclosed in the region of volume $V$ and overdensity $\delta$. Eq.~\ref{eq:ngal_def} would give a standard value as Eq.~\ref{eq:standard_number} due to the marginalizing over the density distribution $p(\delta)$. This distribution is given by the $\Lambda$CDM theory as a Gaussian with the variance equal to $\sigma_M$ on some scale. On the other hand, the usefulness of the conditional mass function is in answering questions about the local environment. If we know the density of the region we are interested in, we can write the conditional number of halos as

\begin{equation}
    N_{\rm halo} (\vec{r}| \delta) = V\cdot \int dM \cdot \frac{\textrm{d}n(>M, z|\delta, M_{\rm vol})}{\textrm{d}M} 
\end{equation}

where $\vec{r}$ indicates that $N_{\rm halo}$ is not a global average, but a spatially varying quantity. This result also extends to conditional number of galaxies:

\begin{equation}
\begin{split}
    N_{\rm gal} (\vec{r}| \delta) = V\cdot \int &  dM \int \textrm{d}M_{\rm UV}\cdot \frac{\textrm{d}n(>M, z|\delta, M_{\rm vol})}{\textrm{d}M} \cdot \\ & p(M_{\rm UV} | M) p(M_{\rm UV}) ,
    \label{eq:uncond}
\end{split}
\end{equation}

In practice, overdensity $\delta$ is not directly observable, but we can infer this quantity using the information from the presence of galaxies. The presence of a bright galaxy is a good indicator of an overdense region \citep{Munoz2008, Lim2025}. This motivates us to condition the number of (generally fainter) galaxies on the observation of a galaxy with given $M_{\rm UV, 0}$, i.e., $p(\delta|M_{\rm UV,0})$, where a bright galaxy acts as an environmental tracer. We write this as 
  
\begin{equation}
\begin{split}
    N_{\rm halo}(\vec{r}|M_{\rm UV, 0}) = V\cdot \int dM \int\textrm{d}\delta & \cdot \frac{\textrm{d}n(>M, z|\delta, M_{\rm vol})}{\textrm{d}M}\\ & \cdot  p(\delta|M_{\rm UV, 0}),
\end{split}
\end{equation}
where instead of marginalizing over $\delta$ with a prior distribution given by a Gaussian, we are marginalizing over the conditional distribution $p(\delta | M_{\rm UV, 0})$.
Again, we need to account for the fact that we are observing galaxies and not halos (i.e., Eq.~\ref{eq:pM_first}). This additional marginalization leaves us with the expression

\begin{equation}
\begin{split}
N_{\rm gal}(\vec{r}|M_{\rm UV, 0})  & = V\cdot \int dM \int\textrm{d} \delta\int\textrm{d}M'_{\rm UV }\cdot \\ & \frac{\textrm{d}n(>M, z|\delta, M_{\rm vol})}{\textrm{d}M}  \cdot p(M'_{\rm UV}|M) p(M'_{\rm UV}) p(\delta | M_{\rm UV, 0}),
    \label{eq:n_gal_conditioned_app}
\end{split}
\end{equation}
where again $p(M_{\rm UV}')$ defines the observational strategy. In the expression above, the only uncertainty is how to write $p(\delta|M_{\rm UV})$.

\subsection{Conditional density distribution}
\label{sec:cdd}
To compute $p(\delta|M_{\rm UV,0})$, we invert the problem using Bayes’ theorem. First we note that the differential probability of finding a halo of mass $M'$ in a region of volume $V'$ is proportional to the normalized halo mass function (which is not the same as observability defined above).

\begin{equation}
    p(M'|V, \delta) \textrm{d}M' = \frac{\textrm{d}n}{\textrm{d}M'}({\delta},V)\textrm{d}M' / N_{\rm gal}(V).
\end{equation}

From which we can use Bayes' theorem to write:

\begin{equation}
    p(M'|\delta) p(\delta) = p(\delta|M')p(M')
    \label{eq:bayes_1}
\end{equation}

In the above, $p(\delta)$ is obtained from the assumption of Gaussian distribution of Eulerian overdensities, while $p(M')$ is obtained from the mean halo mass function. The expression Eq.~\ref{eq:bayes_1} has a simple interpretation. Our prior about the density distribution is given by $\Lambda$CDM prediction that densities are distributed with a Gaussian distribution. Observing a halo in a field alters our knowledge about the density of this region and $p(M'|\delta)$ thus acts as the "likelihood". From here $p(M')$ acts as the Bayesian evidence and $p(\delta|M')$ becomes the posterior we are after.
We can condition over $M_{\rm UV}$ by performing marginalization once more:

\begin{equation}
    p(\delta|M_{\rm UV}) = \int \textrm{d}M' \frac{p(M'|\delta) p(\delta)}{p(M')}p(M'|M_{\rm UV}).
    \label{eq:density_muv_condition}
\end{equation}

where $p(M'|M_{\rm UV})$ can be obtained from $p(M_{\rm UV}|M')$.
We visualize this distribution in the left panel of Fig.~\ref{fig:full_theory_plot} for a region of size $R_{\rm b} = 10$\,cMpc at $z=10.5$.

\subsection{Final expression}
\label{sec:final_expression}
The final expression for the number of neighbors around a bright galaxy is obtained by getting back to Eq.~\ref{eq:n_gal_conditioned} and plugging in Eq.~\ref{eq:density_muv_condition}. The final expression is:

\begin{equation}
\begin{split}
N_{\rm gal}(\vec{r}|M_{\rm UV, 0}) & = V\cdot  \int dM \int\textrm{d} \delta  \int\textrm{d}M_{\rm UV } \  \\ & \times \frac{\overline{\rho}_0}{M} \frac{\textrm{d}f_{\rm coll}(>M, z|\delta, M_{\rm vol})}{\textrm{d}M} \cdot p(M_{\rm UV}|M) \cdot p(M_{\rm UV}) \\ &  \times \int \textrm{d}M' \frac{p(M'|\delta)p(\delta)}{p(M')}p(M'|M_{\rm UV,0})
\label{eq:final_2}
\end{split}
\end{equation}

This quantity is shown in the middle panel of Fig.~\ref{fig:full_theory_plot} for the two models presented in Section~\ref{sec:analytical_framework}.

\section{Conditioning on the total galaxy count}
\label{sec:cond_on_N}
\begin{figure}
    \centering
    \includegraphics[width=1.0\linewidth]{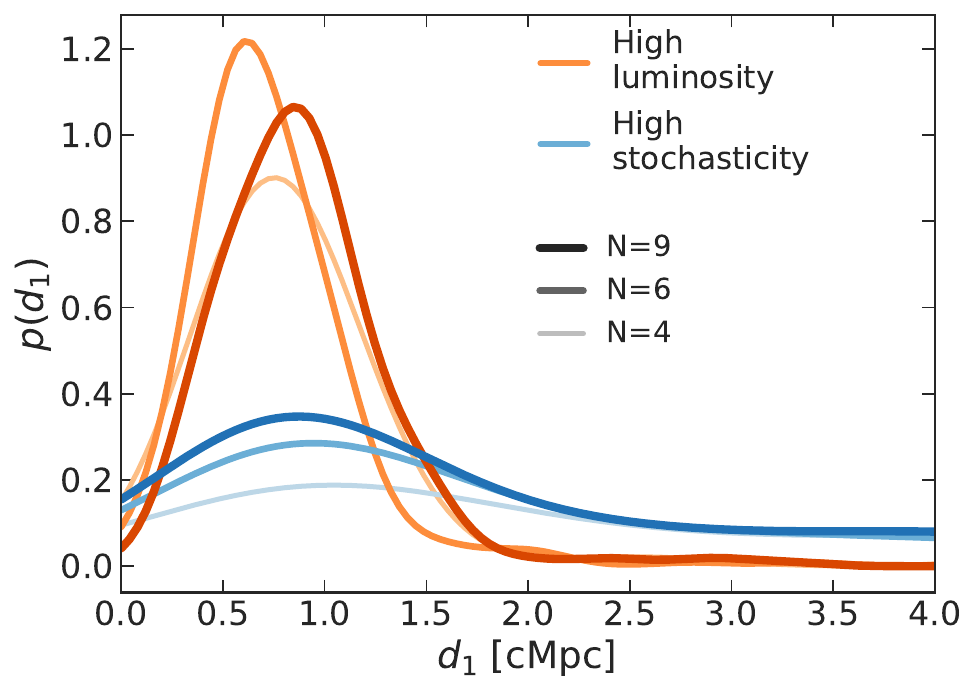}
    \caption{Probability distribution function of the distance to the closest neighbor in $10 \times 10 \times 10$ cMpc$^3$ cutouts for the two models, conditioned on the observation of $N$ neighbors brighter than $M_{\rm UV} = -18.5$.}
    \label{fig:CDF_by_N}
\end{figure}

In Section~\ref{sec:methodology} we showed that the number of fainter neighbors around a bright galaxy is not sufficient to discriminate between extreme models of galaxy evolution, compared to the clustering statistic contained in the distance to the nearest neighbor. However, we can improve the discriminatory power of the framework by conditioning on additional observable properties of the field, rather than relying on a single summary statistic, such as $d_{1}$. As shown in Fig.~\ref{fig:num_dens_sim}, the number of galaxies ($N$) is a useful but insufficient quantity to distinguish between the galaxy models. Here, we therefore condition the distribution of $d_{1}$ on $N$. This approach retains the simplicity and interpretability of a single summary statistic, while robustly incorporating information from all observable galaxies. In practice, we group all simulated cutouts by their total galaxy count and analyze the corresponding distribution of $d_{1}$.

Fig.~\ref{fig:CDF_by_N} shows the distribution of $d_{1}$ conditioned on the total number of galaxies for $N=4$, $6$, and $9$. In the high luminosity model, the variation of $N$ has little impact on the shape of the $d_{1}$ distribution. In contrast, in the high stochasticity model, the distribution becomes flatter as $N$ increases, meaning that close neighbors are less common when fewer galaxies are observed.

The different responses to the conditioning on $N$ can be understood in terms of the halo masses predicted by the two models. In the high luminosity model, the presence of a bright galaxy is a strong indication of an overdense field. As a result, the conditional distribution $p(\delta|M_{\rm UV,0})$ is both shifted to larger values of $\delta$, and relatively narrow (cf. Fig.~\ref{fig:full_theory_plot}). Any further conditioning on the number of galaxies does not significantly modify the inferred distribution. In contrast, in the high stochasticity model, the mapping between UV luminosity and halo mass is weaker, leading to a broader distribution of overdensities for fixed $M_{\rm UV, 0}$. In this case, the total number of galaxies provides an additional constraint on the underlying overdensity of the region. Conditioning on higher $N$ selects more dense environments, pushing the distribution of $d_{1}$ to shorter values. This shows that counting galaxies provides additional information about the underlying density field.

\section{UV luminosity function at $z\in[8,14]$}
\label{sec:app_z_uvlf}
\begin{figure*}[ht!]
    \centering
    \includegraphics[width=\linewidth]{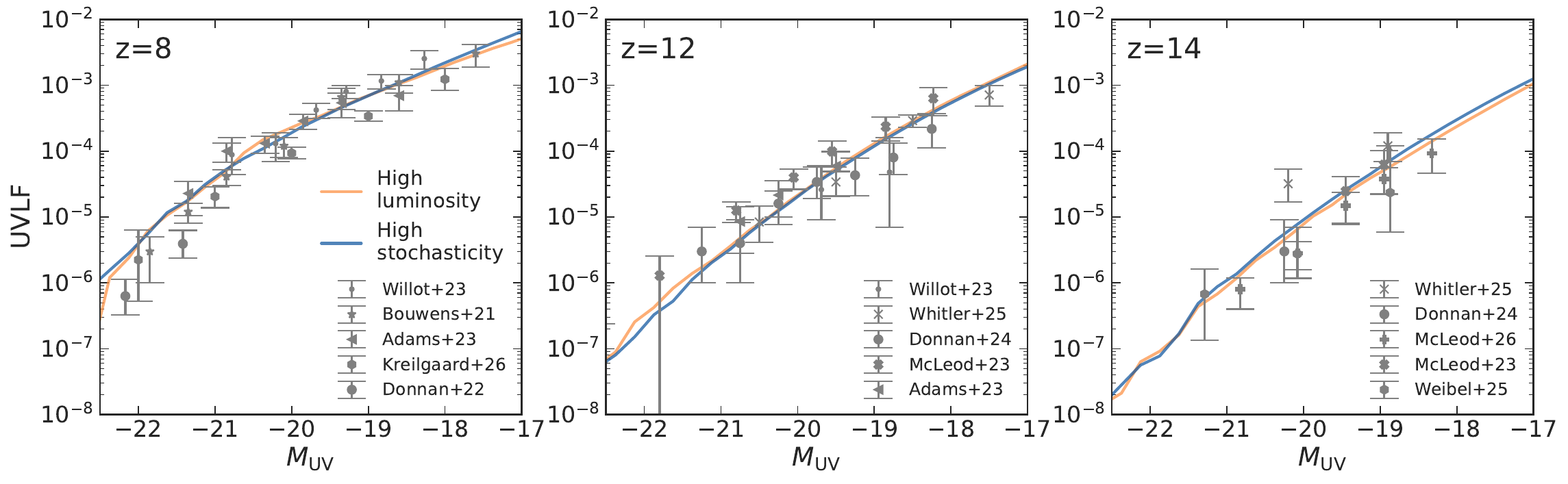}
    \caption{UV luminosity function at different redshifts for two galaxy formation models, one in which the UV magnitudes of all galaxies are intrinsically brighter (orange line), and one in which stochasticity in the $M_{\rm UV} - M_{\rm h}$ connection is significantly increased (blue line). \textit{Left panel:} The UVLF at $z=8$ tuned to match the UVLF observation from \citet{Bouwens2021}, consistent with JWST estimates \citep{Willot2024, Adams2024, Kreilgaard2026}. \textit{Middle panel:} The UVLF at $z=12$ tuned to match the UVLF observation from JWST \citep{Willot2024, Donnan2024, Whitler2025, Adams2024}. \textit{Right panel:} The UVLF at $z=14$ tuned to match the UVLF observation from JWST \citep{Finkelstein2023, Donnan2024, Whitler2025, Robertson2023}.}
    \label{fig:uvlf_highz}
\end{figure*}

In Section~\ref{sec:red_dep} we extended our galaxy formation models, high luminosity and high stochasticity, to other redshifts by adjusting the median evolution to the one presented in \citep{Mason2015, Mason2023}. In this appendix, we present the UV luminosity functions for the resulting models and comparison with observations from HST and JWST \citep{Bouwens2021, Finkelstein2023, Robertson2023, Willot2024, Donnan2024, Adams2024, Whitler2025, Kreilgaard2026}. This is shown in Fig.~\ref{fig:uvlf_highz} for $z\in\{8,12,14\}$. The two models fit the data well for a range of $M_{\rm UV}$ and differences between the models are within the error bars for UVLF estimates for all redshifts.
\end{appendix}

\end{document}